%% file: paper.tex
\documentclass[sigchi]{acmart}% 
\definecolor{RED}{HTML}{ff2e63}
\definecolor{TEAL}{HTML}{08D9D6}
\usepackage[utf8]{inputenc}%
\usepackage{comment}%
\usepackage{amsmath}%
\usepackage{tabularx}%
\usepackage{pgf}%
\usepackage{caption}%
\usepackage{subcaption}%
\usepackage{color}%
\newcommand\todo[1]{{\color{black}{#1}}}%
\newcommand\ok[1]{{\color{black}{#1}}}%
\newcommand\rev[1]{{\color{black}{#1}}}%
\newcommand\transition[1]{{\color{black}{#1}}}%
\usepackage{caption}%
\usepackage{subcaption}%
\usepackage{tikz}%
\usepackage{array}%
\usetikzlibrary{backgrounds,positioning,shapes.geometric, arrows}%
\newtoggle{anonymous}%
\makeatletter
\@ifclasswith{acmart}{anonymous}{
    \settoggle{anonymous}{true}
}{
    \settoggle{anonymous}{false}
}
\makeatother
\copyrightyear{2025}
\acmYear{2025}
\setcopyright{cc}
\acmConference[CHI '25]{CHI Conference on Human Factors in Computing Systems}{April 26-May 1, 2025}{Yokohama, Japan}
\acmBooktitle{CHI Conference on Human Factors in Computing Systems (CHI '25), April 26-May 1, 2025, Yokohama, Japan}\acmDOI{10.1145/3706598.3713556}
\acmISBN{979-8-4007-1394-1/25/04}
\newcolumntype{Y}{>{\centering\arraybackslash}X}%
\newcolumntype{C}[1]{>{\centering\arraybackslash}m{#1}}%
\newcolumntype{L}[1]{>{\raggedright\arraybackslash}m{#1}}%
\usepackage{calc}%
\newcommand{\maxbarwidth}{5cm} % Define the maximum width for the bars.
\newcommand{\barchart}[1]{
    \noindent
    \begin{tikzpicture}
        \pgfmathsetlengthmacro\barlength{#1/1983*\maxbarwidth}
        \definecolor{barcolor}{RGB}{88, 88, 88}
        \fill[barcolor] (0,0) rectangle (\barlength, 0.3cm);
    \end{tikzpicture}
}
\newcommand{\NUMARTICLES}{11,542}%
\begin{document}%
%
% [Citation Practices in HCI]
\title{%
% On the
Past, Present, and Future of Citation Practices in HCI
% \\ in the CHI Community
}%
% \subtitle{A Quantitative Analysis}

\author{Jonas Oppenlaender}%
% \authornote{The author is in charge of his own destiny.}
\email{jonas.oppenlaender@oulu.fi}%
\orcid{0000-0002-2342-1540}%
\affiliation{%
  \institution{University of Oulu}%
  % \zip{90570}%
  \city{Oulu}%
  \country{Finland}%
  % \postcode{40014}
}

\begin{abstract}%
\noindent
Science is a complex system comprised of many scientists who individually make % collective
decisions that, due to the size and nature of the academic system, largely do not affect the system as a whole. However, certain decisions at the meso-level of research communities, such as the Human-Computer Interaction (HCI) community, may result in deep and long-lasting behavioral changes in scientists.
% Academia has been witness to exponential growth in recent years.
% In the field of Human-Computer Interaction (HCI), the field's top conference -- the ACM CHI Conference -- has been expanding
% % , albeit linearly,
% and
%     % The number of papers is exploding exponentially, while peer reviewers are more and more difficult to recruit.
%     the CHI community's quality expectations and standards 
%     % on what is considered a publication-worthy article
%     have risen considerably over the past decades.
% % The growth in publications and the expansion of the research field 
% These developments may lead to behavioral changes in scientists.
% conduct and write up their research.
% Using the ACM CHI Conference as a proxy for the HCI community,
In this article, we provide % quantitative
empirical evidence on how a change in editorial policies introduced at the ACM CHI Conference in 2016 destabilized the CHI research community and launched it on an expansive path,
% We identify clear quantitative changes in citation practices % occurring after the year 2015,
denoted by a year-by-year increase in the mean number of references included in CHI articles.
% CHI authors today are on a different trajectory than before 2015.
% Clearly, the CHI community entered a novel trajectory at this point in time.
% While the growth in references since 2015 is linear, the development is still worrisome as it makes peer review more difficult.
% Based on the data, we trained a linear regression model that predicts the number of references in future CHI articles.
If this near-linear trend continues undisrupted, an article at CHI 2030 will include \textit{on average} almost 130~references.
    % to other articles.
% --- too many to thoroughly vet during rigorous peer review.
%
% We further % dive deep into the CHI community's citation practices and
% present empirical evidence on a number of co-occuring changes since 2016.
    % Our % temporal analysis
    % meta-research provides insights into how the nature and % purpose
    % meaning of citation practices in HCI % 's scholarly work
    % have changed, influenced by factors such as digital accessibility of resources and academic pressures.
    % % Quality vs. Quantity of Citations:
% The observed 
The trend toward more citations % correlates with higher quality research or if it simply
reflects a citation culture where quantity is prioritized over quality,
% This could involve analyzing whether heavily cited papers genuinely contribute to the field's development.
    % This culture % puts enormous strain on the academic system
    % adding to the strain on the academic system
    % and
    contributing to both author and peer reviewer fatigue.
    % putting peer review, as a traditional means of quality control, under pressure.
% We provide recommendations for policy-makers and authors.
Our exploratory analysis % underscores
highlights the
% value of meta-research for research communities and the
profound impact of meso-level policy adjustments on the evolution of scientific fields and disciplines, urging all stakeholders to carefully consider the broader implications of such changes.
    % and their part in the system.
% Future authors are advised to follow the trend to conform to the HCI community's expanding expectations % for maximizing the chances of acceptance
% --- or radically break with the trend and contribute to change in academia.
\end{abstract}%

% \ccsdesc[300]{Human-centered computing~Empirical studies in HCI}
% \ccsdesc[100]{General and reference}
% \ccsdesc[300]{General and reference~General conference proceedings}

\begin{teaserfigure}%
  \includegraphics[width=\textwidth]{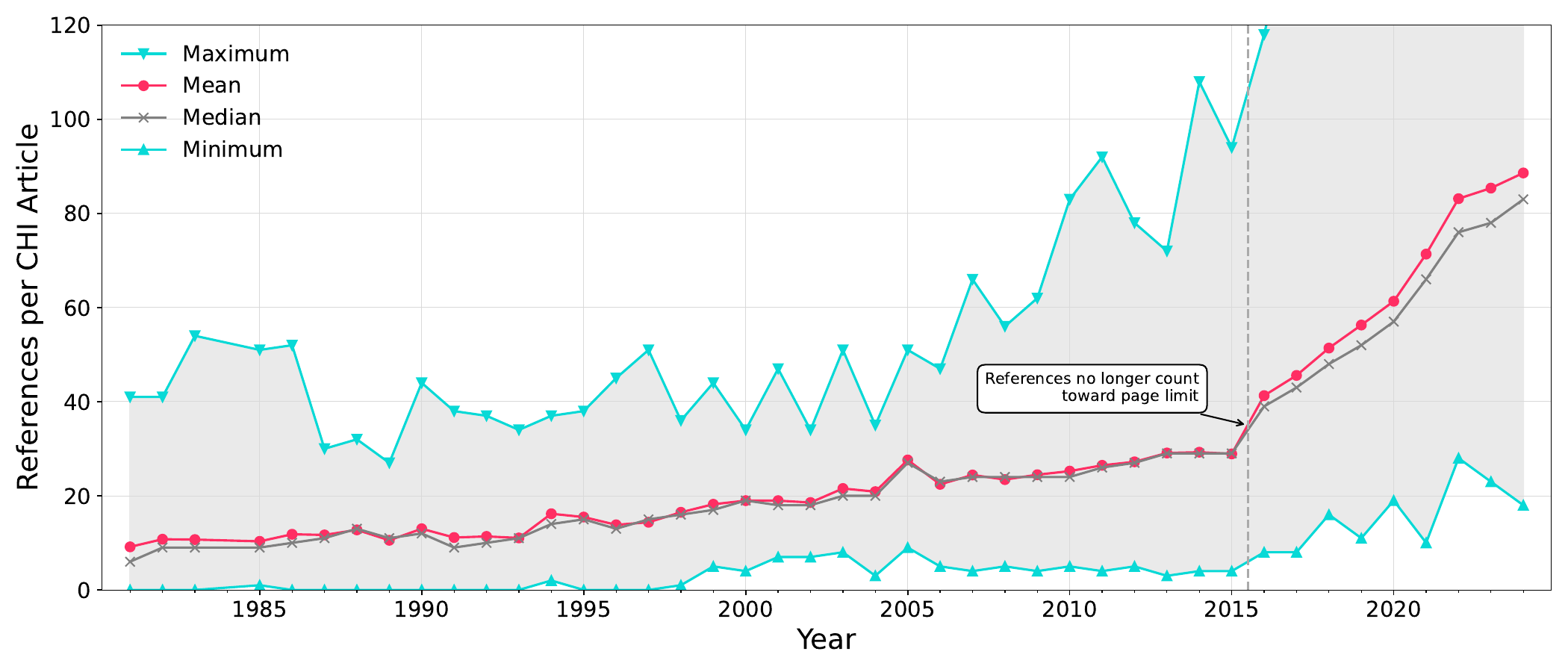}%
  \caption{Number of references per article in the Proceedings of the ACM CHI Conference from 1981 to 2024.}%
  \label{fig:teaser}%
\end{teaserfigure}%
\maketitle%
%%%%%%%%%%%%%%%%%%%%%%%%%%%%%%%%%%%%%%%%%%%%%%%%
%%%%%%%%%%%%%%%%%%%%%%%%%%%%%%%%%%%%%%%%%%%%%%%%
%
%
% \renewcommand*\contentsname{}
% \tableofcontents
%
% ===========================
\section{Introduction}%
\label{sec:introduction}%
% ===========================
% See abstract.
%
% We are living through a time of rapid reconfiguration of society and science.
% With free access to language models ushered in by OpenAI, ... 
%
%     Provide a brief overview of HCI and the importance of citation practices within the field.
%    State the problem or gap your study addresses.
%    Clearly articulate the paper’s main objectives or research questions.
%
Citation serves as the backbone of academic rigor, enabling researchers to build upon previous work, acknowledge contributions, and weave a rich tapestry of interdisciplinary knowledge.
% In the dynamic and evolving field of Human-Computer Interaction (HCI), where the confluence of technology, psychology, design, and ergonomics leads to constant innovation, the role of citations becomes even more critical. 
% Citations not only facilitate the cross-pollination of ideas but also ensure the traceability and credibility of scientific inquiry.
%
% The landscape of academia is undergoing a rapid transformation in the presence of new technologies and unprecedented % exponential
% growth in scientific publications. This expansion, while a testament to the % vibrancy and
% dynamism of the academic field, brings with it a set of challenges that demand % urgent
% attention.
% Among these challenges, the phenomena of author and peer reviewer fatigue emerge as critical concerns, signaling an academic system under strain and in need of thoughtful examination and intervention.%
%
Citation practices, as a tangible expression of scholarly discourse, provide valuable insights into the evolving norms and values of the research community \cite{milestones}.
Editorial policies shape the landscape of academic publishing, influencing not only the structure and content of the research published but also the citation practices of researchers.
Understanding the impact of policy decisions on citation practices is crucial for ensuring that scholarly communication serves both the authors and the broader field effectively.
% Our investigation is propelled by the recognition that a dynamic environment such as the CHI community requires continuous scrutiny to ensure its health and sustainability.
% By focusing on citation practices within the ACM CHI Conference Proceedings from 1981 to 2024, our study offers a lens through which to observe and understand shifts in the community.

In this work, we examine
how the citation practices of researchers in the Human-Computer Interaction (HCI) community have changed after an
    % important significant
editorial policy decision introduced in 2016. In this year, the page restrictions % on references included in articles
at HCI's top conference, the ACM Conference on Human Factors in Computing Systems (CHI), were lifted.
% the impact of an editorial policy decision made in the Human-Computer Interaction (HCI) community. In the year 2016, the page restrictions on articles' reference sections were lifted for articles published at the HCI field's top conference, the ACM Conference on Human Factors in Computing Systems (CHI).
This policy decision had a profound impact on the citation practices of the CHI community.
% We present the results of an exploratory analysis through data visualization and correlation analysis.
\todo{We examine the impact of this policy decision with three research questions.}
We first explore the historic de\-ve\-lopment of the number of references in CHI articles:
\begin{itemize}%
    \item[RQ1:] \textit{How has the number of citations from CHI articles to other works developed in the CHI Proceedings?}%
\end{itemize}%
Our analysis of \NUMARTICLES~articles published at the ACM CHI Conference between 1981 and 2024 provides empirical evidence of a year-by-year increase in the mean number of references included in CHI articles.
This trend was enabled by the change in policies in 2016, allowing authors to include an unlimited number of references in their articles.
The subsequent change in the community's citation practices signifies a departure from previous patterns and suggests the community's entry into a novel trajectory (cf. \autoref{fig:teaser}).
If this trend continues undisrupted, articles at CHI 2030 will include \textit{on average} about 130~references.
Such a scenario poses practical challenges for authors and peer reviewers alike, straining the authors' capacity to % thoroughly
meaningfully engage with the literature and the peer reviewers' capacity to thoroughly vet the cited literature during the peer review process.

We explore a number of potential reasons for the observed increase in the mean number of references included in CHI articles:
% We further explore:
\begin{itemize}%
    \item[RQ2:] \textit{What other factors, beside the 2016 policy decision, could potentially contribute to the observed growth in references, and how have these factors evolved over the years?}
\end{itemize}%
We plot and and visually analyze a number of co-occurring trends.
As part of this, we also investigate whether there is a bias in awarding articles at CHI post-policy change.
% \begin{itemize}
%     \item[RQ3:] \textit{Is there a bias toward awarding high-citation articles in the CHI community?}%
% \end{itemize}%
If articles with a high number of references were to systematically be presented with awards at the CHI Conference, this could send a signal to the community and potentially incentivize authors to include more references in their articles, contributing to the observed growth in the mean number of references per CHI article.
% This research question investigates whether there is a systematic bias in awarded articles that could send a signal to the community and, thus, contribute to the observed growth in the mean number of references per CHI article.

Last, we investigate the significance of these observed trends in shaping the
% mean number of references per CHI article:
expansive citation practices in the CHI community:
\begin{itemize}%
    \item[RQ3:] \textit{Was there a significant change in citation practices of authors at CHI after the editorial policy change in 2016? If so, how do the co-occuring factors contribute to this trend?}
    % \item[H1:] \textit{There was a significant change in the mean number of references per CHI article between 2015 and 2016.}
\end{itemize}%
We investigate this research question with an event study, demonstrating that the policy change at CHI '16 had a profound and destabilizing effect on the CHI community. The CHI community has clearly entered a different trajectory at this point in time (cf. \autoref{fig:teaser}), marked by increasing academic pressures contributing to fatigue of both authors and peer reviewers.

% We examine a number of co-occurring qualitative and quantitative changes in the citation practices of HCI researchers.
% Since 2016, it has become more common for researchers to conduct literature reviews which contributes to the increase in the mean number of references per article.
% The impact of the editorial decision has also been qualitative, with more citations to unrefereed arXiv pre-print articles as well as code and data repositories.

The rising number of references in CHI articles % pattern
% It also
raises % fundamental
important questions about the nature of academic discourse,
the balance between comprehensiveness and focus in scholarly communication,
and 
% raises important questions about
the sustainability of current citation practices. % in the field of HCI.
% and the potential implications for peer review processes.
The challenges brought by the escalating citation practices within the HCI community are multifaceted, encompassing the expansion of the CHI Conference, the diversification of topics and methodologies, and the elevation of standards for what constitutes a publication-worthy contribution.
Such developments, while indicative of progress, also contribute to the increasing complexity of the academic ecosystem, negatively impacting both authors and peer reviewers.
With the volume of references expanding, the feasibility of conducting thorough and effective peer reviews is called into question.
This article discusses a number of potential solutions to this trend.
% A number of potential solutions are discussed to counter this trend.

We argue it is high time for the CHI community to pause, reflect, and 
% suggesting a pivotal moment for reflection and reevaluation 
reevaluate how the HCI field approaches the % foundational
practice of citation.
In light of our findings, this paper aims to spark a conversation within the CHI community and beyond about the future of scholarly communication  and academic publishing. By examining the contextual factors and ramifications of changing citation practices, we invite scholars, practitioners, and policymakers to reflect on past decisions and identify sustainable pathways forward.%
% Whether by conforming to expanding expectations or by advocating for radical change, the choices made by individual authors today will shape the
% % contours of
% academia of tomorrow.
% This study, therefore, not only sheds light on a critical aspect of HCI scholarship but also calls for a radical change in citation practices and a collective reimagining of what it means to contribute to the advancement of science in an era of transformation.
% It is time for the HCI community to reflect on its past decisions, and to identify sustainable pathways forward.
%
%
% Over the last decade, the CHI community has been witness to significant developments, both internally within its community and externally in its interaction with the wider academic and technological landscapes.
%
%
% % ===========================
% \section{The Past}%
% \label{sec:past}%
% % ===========================
% % ===========================
% \section{The Present}%
% \label{sec:present}%
% % ===========================
% % ===========================
% \section{The Future}%
% \label{sec:future}%
% % ===========================
%
%
% ===========================
\section{Background and Related Work}%
\label{sec:background}%
\label{sec:related-work}%
% ===========================
%
% --------------------------------------
\subsection{The Field of Human-Computer Interaction}%
\label{sec:hci-field}%
% --------------------------------------
Human-Computer Interaction is both a field and a discipline \cite{2556288.2556969.pdf,2702613.2732505.pdf}.
    As a field, HCI encompasses diverse disciplines, such as computer science, cognitive science, psychology, design, sociology, anthropology, and more \cite{KIM1995304,2702613.2732505.pdf,2556288.2556969.pdf,oppenlaender2023mapping}. HCI draws from these diverse areas to understand how humans interact with computers and how to design user-friendly systems.
% As a field, HCI includes a broad range of research topics like usability, user experience (UX), interaction design, and accessibility.
% Researchers in HCI investigate how to improve the interactions between users and technology through empirical studies, theoretical frameworks, and applied methodologies.
    As an academic discipline, HCI is taught in structured curricula at many universities in dedicated degree programs, educating students in the principles and practices of HCI.
% These programs often lead to degrees specifically in HCI or related areas like UX design or interaction design.
    Human-Computer Interaction is also recognized as a professional discipline with specific roles, such as UX researchers and interaction designers.
    % usability analysts, and HCI specialists.
% Other researchers have aimed to look past the dichotomy of field versus discipline by framing HCI as problem-solving~\cite{10.1145/2858036.2858283}.
    % most HCI research is about three main types of problem: empirical, conceptual, and constructive. This offers a rich, generative, and 'discipline-free' view of HCI resolves some existing debates about what HCI is or should be. It may also help unify efforts across nominally disparate traditions in empirical research, theory, design, and engineering.
In the remainder of this paper, we refer to HCI as a field.
% Professionals in this discipline apply HCI principles to create and evaluate technology that meets user needs and enhances user satisfaction.
% In summary, HCI is an interdisciplinary field that integrates knowledge from multiple areas to study and improve human-computer interactions, and it is also a distinct academic and professional discipline with specialized education, research, and career paths.

% HCI as an inter-discipline \cite{2702613.2732505.pdf}
% The field of HCI is interdisciplinary \cite{KIM1995304,2702613.2732505.pdf}, drawing from diverse research areas and disciplines, such as user experience, design, computer science, cognitive science, psychology, and many others \cite{2556288.2556969.pdf,oppenlaender2023mapping}.
Each year, a growing number of articles are being published in the field of HCI.
The international top conference in HCI is the ACM Conference on Human Factors in Computing Systems (CHI), which has been held each year since 1981 (with exception of 1984). 
CHI is the most prestigious conference venue for researchers in HCI, with many labs submitting their articles exclusively to only this conference venue.
As such, the CHI Conference can serve as a reference standard for the entire field and discipline of HCI.

From its humble beginnings in the early 1980s, the CHI Conference has been expanding since 2005 in terms of articles accepted at the conference. % (see \autoref{fig:chi-articles}).
    Except for the COVID-19 pandemic years in 2021 and 2022, the CHI Proceedings have exhibited near linear growth since 2005.
    In 2024, the CHI Proceedings exceeded 1,000~articles.
    % for the first time in CHI's existence.
    While this expansion is plan-driven and directed by CHI's Steering Committee, it also reflects the
    % need to capture sub-fields of research in the
    expansion and diversification of the wider field of HCI.
% As the HCI field has grown and evolved, the conference's research scope has broadened and the range of relevant literature has expanded.

% \begin{figure}[!htb]%
%   \centering%
%   \includegraphics[width=\linewidth]{figures/dataset.pdf}%
%   \caption{Number of research articles published in the CHI Proceedings in the years 1981 to 2024 (excluding session details, panel sessions, and abstract-only entries).}%
%   \label{fig:chi-articles}%
% \end{figure}%
\label{sec:chigrowthdiscussion}

% <stress the importance of citing other works for HCI researchers>
In this expansive and evolving research landscape, the practice of citing relevant works is important for HCI researchers.
Citations serve as a vital link, connecting current research efforts with the rich history of past works. Citations acknowledge the contributions of other researchers and weave a rich tapestry of knowledge that informs and enriches new inquiries. For a field as interdisciplinary as HCI, where the integration of diverse perspectives and methodologies is crucial, citations provide a structured way to navigate and summarize the vast array of existing knowledge for readers without deep subject-matter knowledge and expertise.
Further, citations enhance the credibility and rigor of academic discourse, enabling scholars to build upon a verified body of work. Proper citation practices contribute to the integrity of the HCI field, fostering an environment of collaboration and continuous learning. As such, citations are not merely a scholarly obligation but a critical element % that underpins the
in the academic and intellectual growth of the HCI community.

% ------------------------
\subsection{Meta-Research on the CHI Community and the Field of HCI}
% ------------------------
While the majority of HCI researchers conduct research \textit{within} their field, some researchers have also examined aspects \textit{about} the field of HCI.
This is commonly referred to as \textit{meta-research}~\cite{document.pdf}.
Meta-research is valuable, as it enables the research community to reflect on its research practices~\cite{3651965.pdf,meta-HCI}.
The CHI Proceedings, in particular, are a % suitable and
fruitful study subject for meta-scientific and bibliometric inquiries in the field of HCI, due to the size and importance of the CHI Conference for the field of HCI.
\citeauthor{1518701.1518810.pdf}, for instance, presented a scientometric analysis of the CHI Proceedings, focusing on organizations and countries that contribute to CHI \cite{1518701.1518810.pdf}.
Their work highlighted the difficulty of judging quality in the context of best paper awards, finding a mismatch between awarded papers and citations received.
Our paper also includes an analysis of awards, however we focus on analyzing whether there is a systematic bias in awarded papers.
    % to identify whether papers including above-average references are more likely to receive an award.

\citeauthor{CHI_Bibliometrics_20190301.pdf} presented a bibliometric analysis of the CHI Proceedings (from 1981 to 2018), %, excluding 1984), 
using a citation network analysis to identify emerging topics in CHI \cite{CHI_Bibliometrics_20190301.pdf}.
\citeauthor{2556288.2556969.pdf} also conducted a bibliometric analysis of
% the intellectual landscape of the CHI Conference.
CHI publications \cite{2556288.2556969.pdf}.
Their keyword co-word analysis quantified and described the thematic evolution of the HCI field based on 3,152~CHI articles published between 1994 and 2013. Like our work, their article includes a comparison between two time periods (1994--2003, 2004--2013), highlighting the underlying trends % and shifts
in the HCI community.

\citeauthor{3290607.3310429.pdf} presented a quantitative meta-research study on paper writing at CHI, including an investigation on citations received \cite{3290607.3310429.pdf}.
Their analysis of 6,578~CHI papers found that readability, title length, and novelty influence citation counts.
Our work differs in that we do not investigate writing and only focus on citations given, not received.

Another area of meta-research is the use of literature reviews in HCI.
\citeauthor{3544548.3581332.pdf} conducted an analysis of the CHI Proceedings and ACM Transactions on Computer-Human Interaction (TOCHI) including 189~articles \cite{3544548.3581332.pdf}.
Like in our work, the authors note an ``insufficient consensus of what to expect of literature reviews in HCI.''
% Similarly, \citeauthor{3685266.pdf} surveyed systematic literature reviews in HCI and found that ``descriptions of their procedures are often inadequate'' \cite{3685266.pdf}.
The diverse types and methodologies of literature reviews in HCI prompt us to split our analysis of literature reviews into a broad and narrow part.

Finally, \citeauthor{1520340.1520364.pdf} presented a quantitative analysis of CHI, examining author counts, gender, and repeat authorship \cite{1520340.1520364.pdf}. %, motivating questions about what the preferred state of CHI should be.
Among other findings, the work identified a trend toward an increasing number of authors per paper.
We identify the same trend in our work, and similar to \citeauthor{1520340.1520364.pdf}, we aim to encourage a discussion about what the preferred state of CHI should be.%
%
% \citeauthor{10.1145/2858036.2858498} examined sample size reporting in HCI \cite{10.1145/2858036.2858498}, determining local standards for sample size within the CHI community. Their analysis included manuscripts published at CHI '14.
% Related to sample size, several authors investigated the ``WEIRDness'' of participants \cite{henrich2010.pdf,p2425-sturm.pdf}.
%
% types of research contributions in HCI
% \cite{2907069.pdf}
%
%
%
%
% the top aim of researchers is to publish papers
% --> transition to issues with peer review
%
%
%
%

% ===========================
\section{Method}%
\label{sec:method}%
% ===========================
In this paper, we employ both exploratory analysis (with descriptive statistics and visual analysis) and statistical analysis with linear regression models % to predict future values
and an event study specification.
%draws on descriptive statistics and visual analysis.
% The research involved counting and analyzing data collected from the ACM Digital Library, as explained in the following sections.%
\transition{%
% The following sections describe how and what data was gathered (Section \ref{sec:dataset}), how the 
The research involved gathering, quantifying, and analyzing data from the ACM Digital Library (ACM--DL), as detailed in the following section.%
}%
%
% ===========================
\subsection{Data Collection}%
\label{sec:dataset}%
% ===========================
The ACM Conference on Human Factors in Computing Systems (CHI) is the HCI field's largest and most important annual conference.
Therefore, the Proceedings of the CHI Conference are a proxy for investigations into the wider field of HCI.
In particular, the digital proceedings are a fertile ground for investigating the citation practices of HCI authors.
The CHI Proceedings are stored in ACM's Digital Library, which is the main source of data in our research.%

% We scrap\-ed the number of references from the ACM Digital Library (ACM--DL) for each article published in the CHI Proceedings from {1981 -- 2024}.\footnote{With exception of the year 1984 in which the CHI Conference was not scheduled.}
% Session details, panel sessions, keynotes, and abstract-only entries were excluded using regular expressions.
% % For 93~articles, we manually looked up the number of references due to the number of references being capped at 150 in the ACM--DL.
% For {150}~articles, we manually collected the number of references because the ACM Digital Library limits the number of references displayed to a maximum of 150 references.
% We additionally identified and manually corrected six instances where the number of references was erroneously reported as zero in the ACM--DL.
% This resulted in a set of DOI identifiers with their respective number of references.

We collected all articles published in the CHI Proceedings from 1981 to 2024 (with exception of 1984) from the ACM--DL.
We focused on full research articles and excluded articles in the companion proceedings or extended abstracts.
We further excluded session details, panel sessions, keynotes, and abstract-only entries using regular expressions.
% We then downloaded the PDF documents for each DOI from the ACM--DL, resulting in a comprehensive dataset of 11,542 CHI articles.
For each article, we collected
    the full set of references included in the article
    % the number of references, 
    and the set of authors from the ACM--DL.
    % For articles with more than 150~references, we manually collected the number of references due to the ACM--DL's display limit.
    % Additionally, we identified and corrected six instances where the number of references was incorrectly reported as zero.
% This resulted in a comprehensive dataset of DOI identifiers with their respective full set of references and authors.
Further, we downloaded the PDF document and collected the full reference section from the ACM--DL for each article. This resulted in a large dataset of CHI articles ({$N={\NUMARTICLES}$}), with their respective full set of references and authors.

\subsection{Analyzing the Mean Number of References in CHI Articles (RQ1)}
To answer RQ1, we count the number of references in each CHI article.
For each year in the CHI Proceedings, we calculate the mean, median, minimum, and maximum number of references.
Visual inspection yields the insight that a clear change in slope occurred after the year 2015 (cf. \autoref{fig:teaser}).

Simple linear regression analysis was conducted to evaluate the extent to which
    % [independent/predictor variable]
    Year
could predict
    % [dependent/outcome/criterion variable]
    the mean and minimum number of References
% This analysis is split into two equal-sized parts, one before the change in policy (2007--2015) and one
after the policy change (2016--2024).
%
% We fit two linear regression models to the mean reference count series, one before the observed change of slope (i.e., the time series from 1981--2015) and one after (2016--2024).
We report the
    regression equation for each model,
    R\textsuperscript{2} values explaining the variance in the models,
    % statistical significance,
    % standard errors,
    t-values and p-values for the slope coefficient,
    and
    the error metrics MAE (mean absolute error) and MSE (mean squared error). 
    The t-values for slope quantify how many standard errors the estimated slope coefficient ($\beta_1$) is away from zero.
    It is used to test the null hypothesis that the slope is zero, which would imply no linear relationship between the predictor (Year) and the response variable (References).
F-tests are conducted testing the overall significance of the regression models, reporting F-statistic and p-value.
The two linear regression models are then used to predict the mean and minimum number of citations in future % instances of the
CHI proceedings. %, respectively.

\transition{%
The mean number of references per CHI article is the key variable in our work. In the following section, we describe how we visualized and analyzed a number of other contextual factors potentially playing a role in shaping the citation practices of the CHI community.}%

% To examine the differences between the linear regression models fitted to the two distinct periods, we employed a Chow Test, a test to determine whether there is a significant difference in the coefficients between two the linear regressions.
% The Chow test is used to test the null hypothesis that the two timeseries (1981--2015 and 2016--2024) have the same regression coefficients.
% The primary output of the Chow test is the F-statistic and associated p-value.

\subsection{Exploring Other Contextual Factors (RQ2)}
\transition{To answer RQ2, we % visually
explore the historical development of a number of % contextual factors
co-occurring trends.
% , as detailed in the following sections.
}%

\subsubsection{Analyzing bias in awarding articles} % (RQ3)
% In Section \ref{sec:awards},
% To answer RQ2,
We examine whether % there is
a systematic bias can be observed in awarded papers in past proceedings of the CHI Conference.
As award, we consider both ``best paper'' awards and ``honorable mentions.''
We count the references in awarded and unawarded articles at the CHI Conference.
We then statically analyze the difference in the mean number of references in articles that received an award  (M\textsubscript{awarded}) and the mean number of references in articles that did not receive an award (M\textsubscript{regular}) for each proceedings year from 2007 to 2024.
% For each proceedings year,
We report the t-test statistic, p-values, 95\% confidence intervals, level of significance, and effect size (using Cohen's $d$).%

\subsubsection{Collaboration and co-authorship at CHI}
For each article in the dataset, we count the number of authors.
We plot the development of co-authorship over time and visually analyze the correlation of growth trends in authors and publications at CHI.

\subsubsection{Literature reviews}
Literature reviews aim to capture the current state of a research area or field. Given the rising number of scholarly articles (at CHI and globally), we expect that
% both the prevalence and size of literature reviews is growing in the CHI community.
CHI authors are increasingly conducting literature surveys.
% We identified literature reviews by searching for the keywords ``literature review,'' ``literature survey,'' and ``systematic literature review'' in each CHI article's full text (not including the references section).
To study this, we identified literature reviews by searching each CHI article's full text (excluding the references section) for the keywords ``literature review,'' ``literature survey,'' and ``systematic literature review.''
% The reference section was identified by searching the article's full text for the last occurrence of the term ``References.''
%%% The reference section was collected from the ACM--DL.
% We then manually verified the occurrences of the keywords in the article's text, making sure to only include articles that refer to having conducted a literature review.
We manually verified each occurrence, ensuring that only articles explicitly stating having conducted a literature review were included.
% Note that it is not uncommon for CHI authors to refer to the related work section as a literature review, especially in the earlier CHI proceedings.
% Furthermore, some authors report having conducted an ``extensive'' or ``exhaustive'' literature review, but make no mention of the literature review process.
% Our data represents ``literature review'' in a broad sense, whereas the ``systematic literature reviews'' are respresented in a narrow sense, often referring to articles conducting a literature review with more academic rigor.
Note that CHI authors, particularly in earlier proceedings, often referred to the related work section as a literature review. These instances are included in our plots.
Additionally, some authors claim to have conducted an ``extensive'' or ``exhaustive'' literature review without detailing the process.
Our data encompasses ``literature review'' in a broad sense, while ``systematic literature reviews'' are represented more narrowly, typically indicating articles that adhere to a more rigorous scientific process.
We visualize the results, testing our assumption of a rise in the absolute and relative number of literature reviews at CHI.

\subsubsection{Citations to pre-print articles}
The landscape of academic publishing is changing.
We examine how CHI authors cite pre-print articles.
Pre-print articles are versions of scholarly % or scientific
articles that precede publication in peer-reviewed journals and conference venues.
While traditionally shunned upon due to not having undergone formal peer  review, it has become common place in some research fields, such as Machine Learning, to % upload and also
cite pre-print articles.
% Current research in Machine Learning cannot be written about without referring to ArXiv pre-prints.
    % Pre-print archives offer a convenient way for authors to claim priority capital, establishing the authors' precedence in scientific discoveries or novel ideas~\cite{0809.0522.pdf}.
% before an article is accepted.
One of the most popular pre-print archives is arXiv, launched by Cornell University in 1991. In recent years, an exponentially growing number of articles are being submitted to this pre-print archive.
The high number of available pre-prints and the success of pre-prints in other fields may % incentivize
motivate CHI authors to also include more citations to pre-prints in their CHI articles.
    % \todo{thus contributing to an increase the average number of references per article.}

To investigate this, % in more depth, we % scraped
% collected
% the references from the ACM--DL for each article in the CHI Proceedings.
we % then
identified the number of citations to arXiv pre-prints in the references of each article in the CHI Proceedings % (1981--2024)
% published after 1991
by counting the number of references including % at least one 
a mention of ``arxiv.''
    % is mentioned in the articles' reference section.
%%% As above, the reference section was collected from the ACM--DL.
% the reference section was identified by searching the article's full text for the last occurrence of the term ``References.''
For counting the number of occurrences of arXiv in the references section, we used a case-insensitive regular expression that accounts for hyphenation within the word arXiv due to potential line breaks.
We then plot the absolute and relative number of citations to arXiv pre-prints for each proceedings year and visually analyze the results.%

\subsubsection{Citations to datasets and software code repositories}
With the rise of % big data and 
data-driven research methodologies and machine learning, CHI papers may have started referencing more data sources, scientific software and tools, code repositories, and prior data-heavy studies, which could be another potential factor contributing to the increase in the mean number of references in CHI articles.
% These prior studies may be cited for development or validation purposes.
% This contributes to the observed increase in the average number of references per article.
We investigate this trend by counting the number of references to five
    % common
    popular
open data and source code repositories:
\begin{itemize}%
    \item \textit{osf.io}, a cloud-based management solution for open % access
    science, often used to share datasets and other data related to HCI studies,
    \item \textit{zenodo.org}, an open repository for datasets and other research related digital artefacts,
    \item \textit{github.com}, a source code repository that, with git LFS, can also store big datasets,
    \item \textit{kaggle.com}, a web-based platform for sharing datasets for  machine learning, and
    \item \textit{huggingface.com}, a repository for storing and sharing machine learning related files.
\end{itemize}%
% We examine the number of references to these data and code repositories.
As above, we limit our investigation to the articles' reference section.
% which we
% scraped
% collected
% from the ACM--DL.
We ignore citations in footnotes within the article's full text to make the results comparable to the other findings in this article.
%%% As above, we used the last occurrence of the word `References' to identify the references section.
We plot the number of citations to data and source code repositories and visually analyze the results.%
\subsubsection{Citations to questionable publishers}
Predatory journals are questionable publications that accept articles with minimal peer review. % \cite{489179a.pdf}.
These journals are generally regarded as having a lower quality compared to more rigorously peer-reviewed journals \cite{489179a.pdf}.
Without trade-off cost between references in CHI articles, CHI authors may cite articles from potential predatory journals and questionable publishers more often.

To investigate this, we compare the references included in CHI articles with entries in Beall's List of Potential Predatory Journals and Publishers \cite{489179a.pdf}.
We first downloaded Beall's List\footnote{https://beallslist.net} and merged the lists of journals and publishers into one list.
We removed any additional notes and entries that are generic, such as ``Qualitative Research'' and ``American Journal,'' since these entries would produce many false positive matches.
We then checked for exact matches between the entries in Beall's List and the references included in CHI articles (1981--2024), using basic string normalization techniques (e.g., lower-casing, removal of line breaks and special characters, and expansion of ampersand characters into the word 'and').
This comparison yielded 5,126 potential matches between entries in Beall's List and CHI references.
Because the names of predatory journals are often designed to partially match with reputable journals \cite{489179a.pdf}, each of the 5,126~potential matches was manually reviewed to sort out false positives.
We plot the number of citations to potential predatory journals and publishers and visually analyze the results.
% We split this investigation into citations to Frontiers Media and the remainder. 

Note that this approach is limited in several ways, pertaining to the limitations of Beall's List and the string matching approach.
First, Beall's List is subjective and it should not be the only source of quality appraisal.
Second, the list contains entries with notes that the journal or publisher may, in fact, not be predatory. 
Beall's List also mentions several instances where a 
previously reputable journal has been taken over by a predatory publisher.
Third, while an archived version is available, Beall's List is no longer maintained.
Finally, exact string matching is imperfect, although we address false positives with manual review.
The presented figures can be considered a lower boundary, as there may be more matches (especially if including generic titles) that are not identified with the exact matching approach.

\subsection{Analyzing the Significance of the Observed Trend (RQ3)}%
To answer RQ3, 
we analyze the impact of the policy change on the outcome variable
    % $\textrm{References}_t$
(mean number of references in CHI articles) with an orthogonalized event study specification. % with time-series regression.
    An event study is a statistical method used to assess the impact of an event or intervention on an outcome by analyzing changes in the outcome variable before and after the event within a defined time window.
The event of interest, here, is the 2016 policy decision to lift page restrictions on the references included in CHI articles, and the event windows are the equal-sized pre- and post-event periods (2007--2015 and 2016--2024).
% \todo{To achieve this, we use a regression with an interaction term or a binary indicator for the post-event period.}
With the event study, we examine both the significance of the observed change in citation behavior among CHI authors and the effect of different confounders on the mean number of references in CHI articles.
% The model controls for
The rich set of covariates includes for each year $t$:
\begin{itemize}
    \item $\textrm{Authors}_t$:
        mean number of authors per article,
    \item $\textrm{Arxiv}_t$:
        mean number of citations to arXiv pre-prints per article,
    \item $\textrm{Repos}_t$:
        mean number of citations to code and data repositories per article,
    \item $\textrm{Reviews}_t$:
        mean number of literature reviews per article,
        % (including systematic literature reviews),
    \item $\textrm{Predators}_t$:
        mean number of citations to potential predatory journals and questionable publishers per article, and
    \item $\textrm{AwardRatio}_t$:
        the relative prominence of awarded articles compared to unawarded articles, defined as the ratio between the mean number of references ($M$) in awarded and unawarded articles in a given proceedings year $t$:\\
        $\textrm{AwardRatio}_t = M_{\textrm{awarded},t} / M_{\textrm{unawarded},t}$.
\end{itemize}
% These confounders are examined in detail in Section~\ref{sec:context}.
\noindent
The regression model for the event study
    can be written as:
    % is:
% \[
% \resizebox{\textwidth}{!}{
% \begin{split}
% \begin{equation}
% \begin{aligned}
% \begin{equation*}
% \begin{equation*}
% \resizebox{\textwidth}{!}{%
% \cdot
\begin{align*}
    \textrm{References}_t = \beta_0 + 
    \beta_1  \textrm{Post}_t +
    \beta_2  \textrm{AwardRatio}_t +
    \beta_3  \textrm{Authors}_t~+
    \\
    \beta_4  \textrm{Arxiv}_t +
    \beta_5  \textrm{Repos}_t +
    \beta_6  \textrm{Reviews}_t +
    \beta_7  \textrm{Predators}_t~+ 
    \\
    \beta_8  (\textrm{Post}_t \cdot t) +
    \beta_t  t +
    \epsilon_t
\end{align*}
%\end{equation*}
% \end{split}
% \]
% }%
% \end{equation*}
% \end{aligned}
% \end{equation}
with $\epsilon_t$ being an unobserved error term.
$\textrm{Post}_t$ is an indicator variable denoting the time period (1 if $t \ge 2016$ and 0 otherwise).
The parameters $\beta_1$ and $\beta_8$ are the key coefficients of interest in this event study specification.
    Coefficient $\beta_1$ indicates
    % a change in intercept between the pre- and post-event periods (2007--2015 and 2016--2024), whereas 
    a mean difference in References between the pre- and post-event periods.
The interaction term $\beta_8$ % between the time variable $t$ and the event indicator ($\textrm{Post}_t$),
captures a change in slope at the event point ($t=2015.5$).
% , controlling for a rich set of
% following potential confounders
The term $\beta_t$ captures trends that vary linearly over time, ensuring that the model accounts for systematic changes that occur regardless of the intervention. % (e.g., gradual increases in citation counts or publication activity).
        % authors and citations to arXiv per year, respectively.
    % For the other variables, % (absolute counts),
    % the coefficients represent the expected change in $\textrm{References}_t$ for a one-unit increase in the total count per year.
% The coefficient $\beta_t$ captures the average annual linear change in References over time (Year $t$).

The time variable $t$ is centered around its mean (the event point, $t=2015.5$).
Orthogonalization was applied to address multicollinearity among predictors, ensuring that the estimated coefficients reflect independent contributions of each variable.
% To address multicollinearity in the model, we use orthogonalization to remove the shared trend with time by
That is, we regressed each variable on $t$ and use only the residuals $u_t$ for regression:
% \[
\begin{equation*}
    \textrm{Variable}_t = \gamma_0 + \gamma_1 t + u_t 
\end{equation*}
% \]
where $u_t$ is the part of the variable not explained by $t$.
This approach retains variability in the levels of variables
    % which is critical for interpreting intercept and slope changes,
and removes the temporal correlation while keeping other meaningful relationships.
% In the orthogonalized model, each coefficient represents the unique contribution of its corresponding variable to the outcome, independent of all other predictors. For example, the coefficient for Post indicates the change in the intercept after the event, while the coefficient for the interaction term $\mathrm{Post} \cdot t$ reflects the change in the slope post-event, both adjusted for multicollinearity.
%%% All variables are standardized by subtracting the mean and dividing by the standard deviation. Standardizing rescales the variables to have unit variance, making it easier to directly compare the relative effect sizes of different predictors (as their coefficients are on the same scale).
    % Therefore, the coefficients % $\beta_2$--$\beta_7$
    % in the model represent the expected change in $\textrm{References}_t$ for a one-standard-deviation change in the respective covariate.
    %%% Therefore, the coefficients % $\beta_2$--$\beta_8$
    % represent the expected change in $\textrm{References}_t$ for a one-unit increase in the mean of the respective covariates.
% Awards are not part of the event study, because complete information on awarded articles is not available for all CHI proceedings in the ACM--DL.
We report the coefficients, t-statistic for each coefficient with p-values, 95\% confidence intervals, and overall significance of the model with an F-test.

\section{Findings}%
\label{sec:results}%
% ===========================
Over the course of 43~years (from 1981 to 2024, with exception of 1984), authors at the CHI Conference cited a total of 558,080 articles.
The mean number of references per CHI article over the entirety of the CHI Proceedings is 48.35~references ($SD=35.09$, $Min=0$, $Max=496$).
However, the field of HCI has evolved from its humble beginnings, and a more in-depth look at these descriptive statistics is warranted.
In particular, \transition{we % focus on
examine the increase in the mean number of references per CHI article in the following section.}%

% The change could also be a sign of a change in dynamics of the evolving CHI Conference.
% \todo{
% To investigate this theory, we conducted a temporal analysis of cited works. For each CHI proceedings, we analyzed the publication year of cited works. A trend toward citing more recent papers could suggest a rapidly evolving field, while a broad range of publication years might indicate a diverse set of influences.
% }
% \autoref{fig:averageage} depicts the age of citations in each proceedings.

% ======================================
\subsection{The Rising Number of References in CHI Articles (RQ1)}%
\label{sec:general-trend}%
% ======================================
We first provide a historical overview of citation trends in the CHI Proceedings, from the conference's beginnings to the current trend.
% More specifically, we present the historic development of the minimum, mean, and maximum number of references in CHI articles.
We then extrapolate this trend into the future to provide an outlook for the year 2030. This year is not too distant (only five years from the time of writing), and we believe focusing on the year 2030 in our prediction of future trends will -- if the current trend is not broken -- provide an accurate and realistic estimation of the future of citations at CHI.%
\subsubsection{Historical development}%
% ---
\input{FIG-REFS-FULL}
% ---
% ---
\input{FIG-CFPs}%
% ---
In the early years of the CHI Conference, it was not uncommon for authors to submit an article without a single citation to other works. Zero-reference articles continued to be a part of the CHI Proceedings with only few exceptions until the year 1997 (see \autoref{fig:teaser} and \autoref{fig:boxplots}).
From 1998 to 2017, the minimum number of references per article has been positive and in the one-digit range. Since 2017, the minimum number of references has increased into the double-digits.
    %, although this increase takes place at a lower rate than the increases in the mean and median number of references per article.

Between 1981 and 2015, the mean number of references per CHI article showed only modest linear growth (cf. \autoref{fig:teaser} and \autoref{fig:boxplots}).
After 2015, a clear and sudden quantitative change in the CHI community's citation practices took place.
% (see \autoref{fig:teaser} and \autoref{fig:boxplots}).
Clearly, the CHI community entered a novel trajectory at this point in time, characterized by a year-by-year increase in the mean number of references per article.\footnote{Throughout this article, we refer to the mean number of references. As depicted in \autoref{fig:teaser}, the median number of references follows a very similar trend.}
While there were on average 28.8~references per article in the 2015 CHI Proceedings, the number of references per CHI article more than tripled to an average of 88.6~references per CHI article in 2024.

% To highlight the change in trajectory, % and simplify the visualization, we did not include all data points in \autoref{fig:teaser}.
For about two decades since its inception, the maximum number of references per CHI article hovered at around 40. After 2015, the maximum number of references skyrocketed to up to 496 references in one article in 2022.
The box plots in \autoref{fig:boxplots} demonstrate that the maximum number of references per CHI article has become more volatile in recent years.
During the years of the COVID-19 pandemic (2020 and 2021), one can observe a slump in the maximum number and upper quartile of the number of references per CHI article.
However,
there is, on average, a clear tendency for CHI authors to include more references in their articles compared to the years before 2015 and the mean number of references included in CHI articles is monotonously growing each year.%

% The number of references per article between 2006 and 2015 can be approximated with linear regression to a function \ok{$y_{t+1}={m_1}y_t + b_1$, with $m_1=0.79$, $b_1=-1555.92$, and $y_{t} \in [2006, 2015]$, with mean absolute error $MAE=0.50$ and mean squared error $MSE=0.38$.}
% The trend after 2015 is approximated by the function \ok{$y_{t+1}={m_2}y_t + b_2$, with $m_2=6.68$, $c=-13435.27$, and $y_{t}>2015$ (with $MAE=1.88$ and $MSE=6.16$).}

% The Chow Test indicates a substantial divergence in the regression parameters across the two periods under consideration ($F=145.47$, $p=4.30×e^{-10}$).
%  % signaling a highly significant difference between the models.
% This % extremely
% low p-value strongly refutes the null hypothesis, suggesting that the regression parameters (slope and intercept) for the two linear regression models M1 and M2 are significantly distinct.
% This statistical evidence indicates the presence of a notable change in the trend between the two periods \todo{(H1)}, corroborated by a clear shift in the slope.

\subsubsection{Extrapolation of the observed trend}%
% The results of the linear regressions are summarized in \autoref{tab:linear-regression-stats}.
% A significant regression was found (F([df for regression],[df for residual] = [F value], p = [p value]. The R2 was [R2 value], indicating that [independent variable] explained approximately [R2 multiplied by 100]\% of the variance in [dependent variable].  The regression equation was:
% [dependent variable] = [constant] + [slope of the regression line]([independent variable]).
% That is, for each one [independent variable unit of measurement] increase in [independent variable], the predicted [dependent variable] [increased/decreased] by approximately [slope of regression line] [dependent variable unit of measurement].
% Confidence intervals indicated that we can be 95\% certain that the slope to predict [dependent variable] from [independent variable] is between [lower bound of confidence interval for independent variable] and [upper bound of confidence interval for independent variable].%
We extrapolate the observed trend using simple linear regression models (LM\textsubscript{1} and LM\textsubscript{2}; see \autoref{tab:linear-regression-stats}) to predict the mean and minimum number of references per article in future instances of the CHI Conference Proceedings (see \autoref{fig:predictions}).
Both models are significant ($F(1,7) = 308.96$, $p < 1e^{-6}$ and $F(1,7) = 6.81$, $p = 0.03$, respectively), with model LM\textsubscript{1} explaining 98\% of the variance in the mean number of references.
As depicted in \autoref{fig:predictions}, the year-by-year growth in the mean number of references follows a near-linear pattern.
Each year since 2016, the predicted mean number of references in CHI articles increases by approximately 6.45 references.
We estimate that if the current observed trend is not broken, articles published at CHI 2030 will contain on average 129.5~references.
The minimum number of references per article increases by 1.8 references per year.
It is predicted to reach 33.7~references in 2030, which is more than the mean number of references from the year 2015 (28.9~references).%
% ---
\input{TAB-REGRESSION}%
% ---
% ---
\input{FIG-PREDICTION}%
% ---
%
%
% --------------------------------------
\subsubsection{Root cause}%
\label{sec:chi}%
% --------------------------------------
A diverse and complex community consisting of thousands of members, such as the CHI community, is not likely to suddenly change its course, unless change is mandated from top down.
In our case, an editorial policy change % editorial
% Lifting the page restrictions % in 2016
launched the CHI community on its expansive path.
% A sudden change in trajectory like the one observed in \autoref{fig:teaser} is not probable to have a distinct root cause originating from a change in policies, especially in a community as diverse and complex as CHI.
In this section, we discuss the policy change that
is
% can be considered
the
% main explanatory factor
root cause
for the observed non-linearity in the CHI community's trajectory.% trend.
%
% We expect change in a community as large and diverse as CHI to take place gradually.
% It is unlikely that a single actor (other than the CHI committee) can introduce a sudden change in a complex system and community such as the CHI community.
%
% There may have been changes in the editorial policies at the CHI Conference, or other shifts in academic publication norms, encouraging or allowing for more extensive referencing.
%
% % Given the change in trajectory evident in \autoref{fig:teaser}, a change in editorial policies and publication norms could have taken place in these years.
% To investigate this, we downloaded the CHI call for papers (CfP) for the years 2014--2016.
% We also downloaded the CHI submission guide and review guide for these three years.
% To compare the different CfPs and guides, we used diff, a command line tool that allows to compare two files line by line.
% We also reviewed the ACM's News Release Archives\footnote{https://www.acm.org/news-release-archives} to see if there was a change in the publisher's policies in or around the year 2015.

The CHI Conference organizing committee (Program Chairs) in consultation with the Executive Committee of the ACM Special Interest Group on Computer-Human Interaction (SIGCHI) introduced a decisive change in the year 2016. In this year, the conference's Call for Papers (CfP) moved away from counting references toward the overall page limit of CHI articles, as highlighted in the excerpts reproduced in~\autoref{fig:cfp-excerpts}.
Articles published at CHI '16 were %, for the first time in the CHI's history,
allowed to include an unlimited number of references, as long as the article's content fit the 10-page limit, with references no longer counting toward this page limit.
While sub-committees were introduced as another % significant
important change in the year 2016, allowing an unlimited number of references clearly is % one of
the main % contributor
cause
% to
of
the increase in the mean number of references identified in our work.%

Further, the % Submission Guide for CHI Authors
``Guide to a Successful Paper or Note Submission''
at CHI 2016 \cite{2016guide} stressed the importance of citing relevant previous work, explicitly asking \textit{``is prior work adequately reviewed?''}
The 2016 Guide also highlighted the importance of reproducibility of authors work, with \textit{``letting others build on your work''} being regarded as \textit{``the entire purpose of a CHI Paper''} \cite{2016guide}.
These direct quotes from the 2016 Guide were stark reminders of the importance of citing prior work and may have contributed to making the community pay special attention to their references in the year 2016.

These policy changes nudged the community on its expansive path.
\transition{Several other factors, such as systematic bias in awards} presented to articles, may have also played a role, as described in the following section.%
\subsection{Additional Contextual Factors Influencing HCI Citation Practices (RQ2)}%
\label{sec:context}%
% ======================================
The CHI community is vast and highly diverse, and there are many factors that potentially shape and make up the citation practices of the members of the CHI community.
While the editorial policy change introduced in 2016 clearly is the enabler of the rise in the number of references included in CHI articles, in this section we take a step back and % shed light on a number of 
% delve (yes, delve!) into .......
% In the following two sections (Section \ref{sec:Evolution_and_Diversification} and Section \ref{sec:Sources_and_Accessibility}), we take a step back and take a look at the
% broader contextual factors
explore a number of contextual factors
affecting the citation practices at the CHI Conference.
% Evolution and Diversification of Scholarship
% Sources and Accessibility
The six factors we investigate are
a potential bias in awards (Section \ref{sec:awards}),
an increase in the average number of authors (Section \ref{sec:authors}), an increase in the number of literature reviews (Section \ref{sec:litreviews}), an increasing number of citations to arXiv pre-print articles (Section \ref{sec:arxiv}), an increase in citations to data and code repositories (Section \ref{sec:datacitations}), and an increase in the number of citations to potential predatory journals and publishers (Section \ref{sec:predators}).
These six factors are used as confounding variables in our event study specification (see Section \ref{sec:eventstudy}).
\label{sec:notexhaustive}
We do not claim that this list of factors is exhaustive. Beside the investigated changes, other observable or latent factors could contribute to the % observed
trend, as is the case with any complex system under change.
%
%
% ======================================
%\subsubsection{Evolution and Diversification of Scholarship}%
%\label{sec:Evolution_and_Diversification}%
% ======================================
%
% While it is likely that the above policy change is the main reason for the observed trend,
% contribute to the change in the community's citation practices observed in the previous section.
%
% In this section, we take a look at several co-occurring factors related to the evolving citation practices of CHI authors.
%
% --------------------------------------
\subsubsection{Are high-citation articles rewarded in the CHI community?}% (RQ2)
\label{sec:awards}%
% --------------------------------------
Each year, awards are given to a select number of accepted articles at the CHI Conference. Typically, between 14.6\% and 24.6\% of articles are given an award at the CHI Conference (based on awards presented at CHI from 2015--2024).    % 2015: 119 awarded / 484 total | 24.586776859504134%
    % 2016: 115 awarded / 545 total | 21.100917431192663%
    % 2017: 121 awarded / 600 total | 20.166666666666664%
    % 2018: 126 awarded / 665 total | 18.947368421052634%
    % 2019: 148 awarded / 702 total | 21.082621082621085%
    % 2020: 156 awarded / 758 total | 20.58047493403694%
    % 2021: 142 awarded / 746 total | 19.034852546916888%
    % 2022: 128 awarded / 637 total | 20.09419152276295%
    % 2023: 128 awarded / 879 total | 14.562002275312855%
    % 2024: 190 awarded / 1057 total | 17.975402081362347%
% Awards are a quality signal for the community.
These awards are a signal to the community on which articles are to be considered high quality. As a quality signal, awards can motivate authors to change their behavior. If articles with a high number of references were to systematically be presented with awards, this could potentially incentivize CHI authors to include more references in their articles.%
% In this section, we examine whether % there is
% such a systematic bias can be observed in awarded papers in past proceedings.
% We analyze the difference in the mean number of references in articles that received an award at the CHI Conference (M\textsubscript{awarded}) compared to the mean number of references in articles that did not receive an award (M\textsubscript{regular}). As award, we consider both ``best paper'' awards and ``honorable mentions.''

% ---
\input{TAB-AWARDS}
% ---

% \begin{figure*}[!htb]%
%   \centering%
%   \includegraphics[width=\linewidth]{figures/awards.pdf}%
%   \caption{Comparison of the mean number of references in awarded articles (best paper and honorable mentions) versus articles without awards.}%
%   \label{fig:awards}%
% \end{figure*}%

% ---
\input{FIG-AUTHORS1}
% ---
% ---
\input{FIG-AUTHORS2}
% ---

% The results are listed in \autoref{tab:awards}.
\autoref{tab:awards} shows no evidence of a significant systematic bias toward awarding high-reference articles in recent years (2021--2024).
    % m_awarded: 92.9578947368421
    % m_regular: 87.65051903114187
    % t-statistic: 1.795037535750761
    % p-value: 0.07293386541635023
    For instance, in the 2024 CHI Proceedings, the mean number of references for awarded articles was 92.96 compared to 87.65 for articles without award.
    This difference is not significant ($p>0.05$).
% That means in recent years, a high number of references is not considered a statistically significant signal of high quality in the CHI community.
%
However, awarded articles had significantly more references than unawarded articles ($p<0.05$; Cohen’s $d = 0.23\ldots0.37$) in the years immediately before and after the change (2014--2020, with exception of 2016).
While awards were also given to high-reference articles in some earlier periods (2007--2010), awards were systematically given to high-reference articles around the policy change (with exception of 2016).
% and this may have contributed to the observed change in citing behavior of CHI authors.%
This systematic awarding of articles with above-average references may have sent a quality signal to authors, thus potentially contributing to launching the CHI community on its current expansive trajectory.%
%%% MOVED TO DISCUSSION:
% The systematic award of articles with above-average references
% % right after the change in editorial policy
% may have sent a quality signal to authors, thus potentially contributing to launching the CHI community on its current expansive trajectory.
% \transition{In the following section, we visually explore the development of five further contextual factors.}%

% --------------------------------------
\subsubsection{Collaborative research is increasing}% T5
\label{sec:authors}%
% --------------------------------------
There has been an increase in collaboration in the field of HCI leading to articles with a greater number of co-authors.
    % \cite{1520340.1520364.pdf}.
This is also reflected in the CHI Proceedings.
\autoref{fig:authors-growth} depicts the number of authors per CHI article for each CHI proceedings year.
The mean number of authors per CHI article has increased over time, from an average of 1.85~authors per article in 1981 to an average of 5.17~authors per article in 2024.
\autoref{fig:authors-growth} also demonstrates that in the years between 2020 and 2024, there have been articles surpassing 20 authors in the CHI proceedings.
The CHI 2024 Proceedings contain an article with 36 authors.
 % articles with a very large number of authors ($>20$~authors) are still rare at CHI, although there have been instances of such articles} in recent proceedings (2021--2024).%

\autoref{fig:refs-per-author} plots the average number of references per author
in CHI articles since the inception of the CHI Conference.
In 1981, a CHI article included on average 5.6~references per author.
This number stayed % surprisingly
stable over three decades, with only a slight increase up to 2015.
But there is a noticeable change in the mean number of references per author after 2015.
Since 2016, spurred by the increase in the mean number of references per CHI article, the mean number of references per author has increased.
In 2024, the mean number of references per author has reached 21.4~references per author.
% We note that while the mean number of references per author has seen a slow increase over time, there is a significant change in trajectory in the mean number of references per author after 2015.
% % Therefore, the change in the average number of references per article occurring after the year 2015 is unlikely to be explained by changes in the number of authors per article.

\autoref{fig:authorcitationcomparison} plots the mean number of authors (right axis) against the mean number of references (left axis) per CHI article.
While there clearly is a non-linearity visible in the latter time series, the former time series has been growing fairly linearly.
The plot visually demonstrates that the growth in co-authorship cannot fully explain the growth in the mean number of references per CHI article.%
% The growth in the mean number of references can therefore not be explained by the growth in the number of authors.
%
%
%
% Figure 7 illustrates the development of the average number of references per author since the inception of CHI. In 1981, a CHI article included an average of 5.6~references per author. By 2024, this number had risen to 21.4 references per author.
% While the average number of references per author has gradually increased over time, there is a noticeable shift in the trajectory of this growth after 2015.
%
%
% --------------------------------------
% \subsubsection{Technological Advancements}
% % --------------------------------------
% In the past decade, there was a significant growth and acceleration in technology-related research areas (such as AI, VR, and UX/UI design). These advancements could have led to a richer body of research for authors to reference in these trending areas.
% We investigate this theory by categorizing the articles into their main subject and plotting these subjects separately in \autoref{fig:categories}.
%
%
%
%
% ---
\input{FIG-LITREVIEWS}
% ---
% ---
\input{FIG-ARXIV}
% ---
% ---
\input{FIG-REPOS}
% ---
% --------------------------------------
\subsubsection{Literature reviews are becoming more common at CHI}%
\label{sec:litreviews}%
% --------------------------------------
In total, we identified 615~papers reporting having conducted a ``literature review'' and 60~systematic literature reviews in the CHI Proceedings from 1981--2024.
The time series in \autoref{fig:litreviews:a} showcases that since lifting the page restrictions in 2015, it has become increasingly common for CHI authors to conduct and publish literature reviews.
    % or basing their design decisions on literature reviews.
About 10\% of all articles in the latest four CHI Proceedings mention having conducted a ``literature review,''
% The relative amount of articles in the CHI Proceedings which mention having conducted a ``literature review'' was about 10\% in the years 2021, 2022, and 2023
with only a slight drop in 2024 (see \autoref{fig:litreviews:b}).
% In this year, we find 90~instances of authors reporting literature reviews.
% However, the scientific rigor of these `literature reviews' varies greatly.
% Further, some authors used ``literature review'' as headline for their related work section, especially in the early proceedings.
% This explains some of the observations in \autoref{fig:litreviews}.
% Note that this development also includes instances of authors referring to their related work section as literature reviews.
% To get a better picture of the actual extent of literature reviews in CHI, we searched for occurrences of the term ``systematic literature review'' in the full text of each CHI article. As above, we manually verified each occurrence of this term.
% Because many authors often do not comment on the thoroughness of their review process, we searched for ``systematic literature review'' in the article texts to identify more rigorous literature reviews.
% From this search and manual verification,

Figures \ref{fig:litreviews:a} and \ref{fig:litreviews:b} demonstrate that systematic literature reviews in CHI are a recent and growing phenomenon.
Before 2016, there was no instance of authors calling their work a systematic literature review. In recent years, it has become increasingly common for CHI authors to conduct systematic literature reviews, with up to 2\% of the published articles in a given proceedings series reporting systematic literature reviews.
In 2024, the CHI Proceedings included 17~systematic reviews (1.6\% of the articles published in that year).%

% ======================================
% \subsubsection{Sources and Accessibility}%
% \label{sec:Sources_and_Accessibility}%
% ======================================

% ---
\input{FIG-PREDATORS}
% ---

% --------------------------------------
\subsubsection{Pre-print articles are increasingly being cited by CHI authors}%
\label{sec:arxiv}%
% --------------------------------------
Even though arXiv was made available to the World Wide Web in 1993, we find that none of the CHI articles before the year 2006 included formal citations to arXiv pre-print articles. % in the articles' reference section.
\autoref{fig:arxiv} depicts the number of citations to arXiv pre-prints in CHI articles since 2006.
We note there is a change in this time series after 2015.
While the mean number of citations to arXiv pre-print articles has only slightly increased since 2016 (from 0.03 references in 2015 to 4.24 references in 2024), citations to pre-print articles have overall become more accepted among some CHI authors.
One article in CHI '24 included 55~references to arXiv.
\autoref{fig:arxivnum} depicts the relative number of articles citing arXiv pre-prints $n$ times, with $n$ split into buckets
(1 citation, 2--10 citations and more than 10 citations). %, with $n$ split into buckets between 1 and 25 citations to arXiv per CHI article.
    From this figure, one can note an increase in the number of articles that cite a high amount of arXiv pre-prints (>10 citations).
    In recent proceedings, over a third of the CHI papers include more than one but less than 10~citation to arXiv.
Given that this change coincides with the year 2016, citations to arXiv articles are one potential factor contributing to the increase in references per CHI article.

% --------------------------------------
\subsubsection{Data-driven research is on the rise at the CHI Conference}%
\label{sec:datacitations}%
% --------------------------------------
\autoref{fig:datarepos1} plots the mean number of references to code and data repositories in CHI articles. While the number of formal citations to such repositories per CHI article is still low, we identify a clear increase in the number of citations to code and data repositories occurring after the year 2015. Before 2015, almost no authors formally cited these repositories, whereas today, such citations have become more common.
In the proceedings of CHI 2024, 147 articles (13.9\%) cite at least one data and code repository (see \autoref{fig:datarepos2}).
    Zenodo and Huggingface were cited too infrequently to include in figures \ref{fig:datarepos1} and \ref{fig:datarepos2}.
    GitHub was cited most often with \todo{115} articles (10.8\%) in the CHI 2024 Proceedings including at least one reference to this repository.
This empirical trend in the citation practices reflects the broader trend toward applying data-driven methodologies and machine learning in HCI research.
% As with arXiv, it also reflects the impact of the decision to lift restrictions on the 

% % --------------------------------------
% \paragraph{Digital accessibility via Open Access}% T3
% \label{sec:openaccess}%
% % --------------------------------------
% \todo{
% By 2015, digital libraries and online databases had become more accessible, allowing researchers to access a wider range of articles more easily. This may also contribute to the observed increase in the mean number of references cited in CHI articles.
% }

% \todo{
% We investigate this by querying unpaywall, an open database of open access (OA) publications.
% For each citation made in a CHI article, we query Unpaywall for the citations' OA status.
% }

% \begin{figure}[!htb]%
%   \centering%
%   \includegraphics[width=\linewidth]{figures/openaccess.pdf}%
%   \caption{\todo{Number of open access articles cited in the CHI Proceedings from 1981 to 2023.}}%
%   \label{fig:openaccess}%
% \end{figure}%

% --------------------------------------
\subsubsection{CHI authors increasingly cite questionable publishers.}%
\label{sec:predators}%
% --------------------------------------
We identified 3,298 citations from CHI articles to potential predatory journals and publishers in the CHI Proceedings (1981 -- 2024).
There is a marked increase in citations to potential predatory journals and questionable publishers since 2016, both in absolute and relative numbers (see \autoref{fig:beall}a and \autoref{fig:beall}b, respectively).

Citations to Frontiers Media -- the publisher which successfully pressured Jeffrey Beall to shut down his list -- account for the vast majority of these citations.
Splitting Frontiers from the rest of potential predatory publications, we can see that citations from CHI articles to journals published by Frontiers Media have monotonously increased since 2015, both in absolute and relative terms (see \autoref{fig:beall}e and \autoref{fig:beall}f).
% Around 0.8\% of all citations in the CHI 2024 Proceedings point to a Frontiers journal (cf. \autoref{fig:beall}f).
In 2024, every CHI paper includes on average about 0.8 references to a Frontiers journal (cf. \autoref{fig:beall}f).
The journals from Frontiers Media cited most often in the CHI Proceedings are 
    Frontiers in Psychology	($n=873$, 39.72\% of all citations to Frontiers Media),
    Frontiers in Human Neuroscience	($n=248$, 11.28\%),
    Frontiers in Robotics and AI	($n=213$, 9.69\%),
    Frontiers in Virtual Reality	($n=130$, 5.91\%),
    Frontiers in Neuroscience	    ($n=86$,  3.91\%),
    Frontiers in Psychiatry	        ($n=68$,  3.09\%),
    Frontiers in Public Health	    ($n=67$,  3.05\%),
    Frontiers in Computer Science	($n=55$,  2.5\%),
    Frontiers in ICT	            ($n=37$,  1.68\%),
    Frontiers in Neurology	        ($n=32$,  1.46\%),
    Frontiers in Artificial Intelligence ($n=31$, 1.41\%),
    and
    Frontiers in Behavioral Neuroscience ($n=26$, 1.18\%).
This is a testament that the CHI community has embraced Frontiers Media as a legitimate publisher.

% ---
\input{TAB-EVENTSTUDY}

% ---

Besides Frontiers Media, the remaining matches are mostly one-off citations to potential predatory journals and publishers.
% (cf. \autoref{fig:beall}c).
While there is an increase in the absolute number of citations to these potential predatory journals (other than Frontiers; see \autoref{fig:beall}c), the relative number of citations remained stable and limited in the recent decade (\autoref{fig:beall}d).
Among these citations, IOS Press ($n=418$) and IGI Global ($n=195$) are occurring most often, followed by
    IntechOpen ($n=36$),
    International Journal of Computer Applications ($n=28$),
    Scientific World ($n=15$), 
    Journal of Physical Therapy Science ($n=13$),
    Cambridge Scholars Publishing ($n=9$),
    Global Media Journal ($n=8$), and
    Research and Reviews ($n=8$).%
% World Applied Sciences Journal ($n=6$),
% Turkish Online Journal of Educational Technology ($n=6$),
% Actapress 6, 
% International Journal of Computer Science and Network Security ($n=6$), International Journal of Information and Education Technology ($n=6$), European Scientific Journal ($n=5$)
%
%
%
%
%
% ===========================
\subsection{Factors Explaining the % Citation Practices of Authors
Rise in the Mean Number of References
(RQ3)}%
\label{sec:eventstudy}%
% ===========================
% The results of the event study are listed in \autoref{tab:eventstudy}.
\noindent
Our event study provides insights into the significance of the change observed in this work and the factors influencing the mean number of references in CHI articles across the study period (see \autoref{tab:eventstudy}).
The indicator variable for the post-event period ($\beta_1$) shows a statistically significant effect ($p = 0.006$), suggesting a notable shift in the mean number of references after the policy change. % (event of interest).
The interaction term ($\beta_8$) is also significant, indicating a significant change of slope post-event.

Among the other % control
    variables,
the awards given at the CHI Conference to high-reference articles ($\beta_2)$ had no statistically significant effect on the mean number of references in articles.
As visually demonstrated in \autoref{fig:authorcitationcomparison},
the mean number of authors ($\beta_3$) also had no significant effect on the mean number of references.
Among the other coefficients,
citations to potential predatory publishers ($\beta_7$) is statistically significant ($p=0.007$), indicating that changes in predatory citation patterns may play a meaningful role in shaping citation practices among CHI authors.
    This can be largely attributed to the publisher Frontiers Media becoming a mainstay in the references of CHI articles since 2015.
Conversely, the coefficients for citations to arXiv ($\beta_4$) and repositories ($\beta_5$), as well as literature reviews ($\beta_6$) are not statistically significant, suggesting these factors may have a limited direct impact on the mean number of references in CHI articles.
The time trend coefficient ($\beta_t$) is statistically significant, pointing to other potential underlying effects explaining the trend in the mean number of references per article.
These findings highlight the complexity of citation dynamics and the multifaceted factors influencing scholarly referencing practices.
As mentioned in Section~\ref{sec:notexhaustive}, this investigation is not exhaustive and there could be multiple other % underlying
potential relationships that warrant exploration in future work.%
%
%
%
% Analysis of Cook's distances for the observations indicates that the latest CHI Proceedings contribute most to the observed trend (
%
%
% \todo{%
% We assert that the observed variations in the data on the mean number of references in CHI articles are not merely coincidental but indicative of a significant shift in the underlying dynamics across the two periods examined (see $\beta_1$, $p=0.016$).
% % ChatGPT nightmare text: This conclusion is bolstered by the rigorous statistical testing employed, providing a robust basis for further investigation into the factors driving these changes.
% }%
%
%
%
%
% ===========================
\section{Discussion}%
\label{sec:Discussion}%
% ===========================
%
%
% It is obvious that the CHI community has changed over the years. Our work paints a clear picture on how several phenomena have co-occurred and shaped  the CHI community's citation practices.
%
% Scientific production is subject to exponential progress. This will, eventually, force the academic community to change their habits and practices. This change will likely be a gradual change.
% However, in our work, we find that change in an academic community as large as CHI can also take place suddenly and, most likely, without the members of the CHI community being even aware of the existence of this change, its origins, and its extent.
%
This article presented meta-research in the field of HCI, providing clear descriptive evidence of quantitative and qualitative changes in the citation practices of the CHI community co-occurring since the year 2016.
Specifically, we have described seven co-occurring changes in citation practices:%
% a clear change in the following \todo{four} factors:
\begin{enumerate}%
    \item
    \textit{An increase in the mean number of references included in CHI articles} (Section \ref{sec:general-trend}):\\
    In the 2016 Call for Papers, references were no longer included in the page limit of CHI articles. This editorial decision launched the CHI community on an expansive path, characterized by a year-by-year increase in the mean number of references included in CHI articles.
    % The mean number of references in CHI articles was used in our event study specification as dependent variable.
    Our event study specification provides empirical evidence of a significant change in citation behavior, with the CHI community departing from established patterns in between the years 2015 and 2016.
    The top-down policy change opened the floodgates, allowing authors to cite more and more works in their articles.

    \item
    \textit{Systematic bias in awarding articles post-policy change (Section \ref{sec:awards}):}\\
    A systematic bias in awarding articles was observed. 
    In the years following the policy change (with exception of 2016), articles with above-average number of references were systematically presented with awards.
    However, the event study found this systematic bias in awarding articles not to be a statistically significant factor in predicting the mean number of references in CHI articles.%

    \item
    \textit{An increase in collaboration and the mean number of references per author} (Section \ref{sec:authors}):\\
    It is becoming more common for authors in HCI to collaborate and the mean number of authors per CHI article is expanding.
    An increase in the number of co-authors means that the set of authors may potentially contain authors from different disciplines. This could imply that authors potentially draw from a broader range of interdisciplinary sources, hence leading to more citations to articles from different fields.
    However, both our visual and statistical analysis found
    no evidence of growth in authorship being a significant factor contributing to the growth in the mean number of references in CHI articles, as also evident in \autoref{fig:refs-per-author}.
    % that the increase in the mean number of authors does not correlate with the observed increase in the mean number of references after the year 2015, as evident in \autoref{fig:refs-per-author}.
    The growth in the mean number of references exceeds the growth in authors (which has been quite linear).
    Author growth alone, therefore, does not serve as explanation for the increase in the mean number of references per CHI article.
    
    \item
    \textit{An increase in the number of literature reviews} (Section \ref{sec:litreviews}):\\
    The corpus of scholarly literature is expanding at a rapid rate, and we found that CHI authors increasingly conduct both systematic and unsystematic literature reviews.
    The editorial decision to lift the page restrictions in 2016 greatly contributed to the growing popularity of literature reviews at CHI.
    The years during the COVID-19 pandemic saw an acceleration of this trend, denoted by an increase in the relative number of CHI articles conducting literature reviews. This can be explained by authors not being able to conduct in-person user studies, a common method in HCI, during the pandemic.
    Literature reviews typically contain an above average number of references,
    % , and systematic literature reviews are becoming more common in HCI.
    % Literature reviews, thus, may be a contributing factor to the increase in the mean number of references per article.
    % To thoroughly cover related work, literature reviews need to cite works from an ever expanding corpus of scholarly literature,
    which may contribute to the increase in the mean number of references included in CHI articles.
    However, as depicted in \autoref{fig:litreviews:a}, the trend toward literature reviews already started before the pandemic.
    The event study found that literature reviews do not have a statistically significant influence on the mean number of references. This can be explained by literature reviews, while growing in popularity, still being relatively rare in the CHI Proceedings. 
    % Therefore, It is likely that the trend of conducting literature reviews contributes to the expansion of the mean number of references per article.
    
    \item
    \textit{An increase in the number of citations to arXiv pre-print articles} (Section \ref{sec:arxiv}):\\
    Entire fields of science rely on arXiv as a means for article dissemination. The field of HCI is not immune to this trend, and it is becoming increasingly more common for CHI authors to cite unrefereed pre-prints.
    Before 2015, citations to arXiv were shunned upon by CHI authors. From 2016 onward, citations to arXiv pre-prints were no longer competing for space with citations to peer-reviewed articles.
    % The editorial decision of lifting page restrictions in CHI articles
    %has removed the trade-off cost between citations to pre-print and published articles, which contributes to the shift in citation behavior.
    The decision to lift the page restrictions at CHI paved the way for this shift in citation behavior.
    Today, many CHI authors cite arXiv pre-print articles. This can also lead to adverse problems, such as pre-prints being cited instead of the published articles.
    While the trend of citing arXiv pre-prints was not statistically significant in the event study, 
    the mean number of citations to pre-prints approaches significance ($p=0.06$), hinting at a potential relationship that
    could develop significance in the future.
    % warrants exploration in future work.
    
    \item
    \textit{An increase in the number of citations to open data and code repositories} (Section \ref{sec:datacitations}):\\
    Before 2015, there were virtually no formal citations to code or data repositories to be found in the reference sections of CHI articles.
    From our analysis, it can be concluded that lifting the page restrictions motivated authors to include more citations to such repositories in their CHI articles.
    From an economic lens, without a page limit, there no longer is a marginal cost to including formal citation to % arXiv or
    code and data repositories, and authors no longer have to trade-off between citations.
    The event study found this trend not to be statistically significant, which could be explained with the relative sparsity of citations to code and data repositories in the CHI Proceedings.%

    \item
    \textit{An increase in the number of citations to predatory journals and questionable publishers} (Section \ref{sec:predators}):\\
    Without a trade-off cost between citations, authors are increasingly including citations to potential predatory journals and questionable publishers in their CHI articles. %, as listed by Beall's List.
    The long-tailed distribution of citations to publishers on Beall's List in \autoref{fig:beall} highlights that the CHI community, overall, has not lost its sense for quality.
    The relative number of citations to smaller potential predatory journals and publishers is negligible compared to the overall number of citations in each proceedings year.
    However, the picture is different with larger questionable publishers (Frontiers Media, IOS Press, IGI Global, and IntechOpen) which are increasingly being cited by CHI authors.
    Perhaps this is yet another expression of the CHI community's obsession with quantity.
\end{enumerate}%

Together, these co-occurring trends shape the citation practices of the CHI community in the shifting landscape of HCI.
The trends reflect an interesting gradual change of pace in academic working in the CHI community starting with CHI '16.
% \cite{2556288.2556969.pdf}.
%(cf. with \citeauthor{2556288.2556969.pdf}'s study from 2014 \cite{2556288.2556969.pdf}).
The trends are also a reflection of broader changes that affect not only the field of HCI, but academia in general \cite{edwards-roy-2017-academic-research-in-the-21st-century-maintaining-scientific-integrity-in-a-climate-of-perverse.pdf}.%

% Many factors could influence how members of the CHI community cite other authors, including simple explanations, such as herd mentality.

Like a frog slowly boiling in water, the CHI community 
is % unwittingly
heating up under the relentless growth of references,
risking a culture where quantity overshadows meaningful discourse and genuine innovation.
A common proverb states: \textit{``things need to get a lot worse before they get better''.}
One has to wonder: At what point will the number of references per CHI article become unbearable for authors and peer reviewers -- at 100 references, 130 references, 200 references?
The pre-2016 limit on the number of pages in CHI articles constituted a ``Schelling fence''~\cite{schelling}~-- a hard limit on the number of pages that could be included in CHI articles (including references).
    Authors could not cross this limit without risking a desk-reject of their paper.
Thus, the space for references and content had to be weighed against each other, and each additional reference reduced the space available for the article's content.
Without these clear boundaries and trade-offs, the CHI community is on a slippery slope~\cite{schelling}, a gradual erosion -- or evolution, if one prefers to look at it this way -- of norms and principles.
The expansive path incentivizes authors to, year-by-year, value quantity over quality and include more references in their articles.
% Like a frog in a boiling kettle, the CHI community will reach its limits at some point, and 

% \todo{%
% It is easy to dismiss the rising trend in the mean number of references as mere numerology or an artifact of general growth in articles.
% However, our citation analysis is conducted without making normative assumptions about whether authors cite works that influence them.
% }%

% The CHI community needs to discuss 
% establishing clear boundaries (Schelling fences) to prevent gradual erosion of norms or principles (slippery slopes).

    One could argue that the observed trend in \autoref{fig:teaser} is simply an effect of the overall growth of publications in the HCI field.
    However, while overall publication growth may certainly be a contributing factor, the growth of the CHI Proceedings cannot fully explain the observed growth in the mean number of references for two reasons.
First, the growth of the CHI Conference Proceedings precedes the observed trend by 10 years, as discussed in Section~\ref{sec:chigrowthdiscussion}.
Second, the CHI Proceedings deviated from linear growth during the COVID-19 pandemic, while the mean number of references in CHI papers still increased steadily and fairly
linearly.
    % , as evident in \autoref{fig:chi-articles}.
    % \todo{Further, there have been years without growth in the number of articles in the CHI Proceedings, yet the increase in the mean number of references is steady and linear.}
Therefore, the increase in the mean number of references per CHI article is likely not a mere side-effect of the proceedings growth at CHI.

What is striking about the observed trend in
% \autoref{fig:teaser} and
\autoref{fig:predictions} is that it follows a relatively stable linear growth pattern.
This near-linear growth takes place in the presence of exponential growth in aca\-dem\-ia, such as the exponentially growing number of scholarly articles.
    % , the increasing number of papers uploaded to arXiv, and the increasing number of people in academia, including students, faculty, and administrative staff.
Future work could investigate why the observed year-by-year growth in references is linear.

This near-linear growth could, perhaps, be explained with the simple psychological models of herd mentality and an availability heuristic. Each year, members of the CHI community observe --- whether consciously or sub-consciously --- an increase in the mean number of references per CHI article.
In other fields of science, evidence has been found that longer articles are cited more \cite{2212.06574.pdf}. Authors may, thus, decide to also include more references in their articles.
% Another, equally simple, explanation could be an availability heuristic where citations are driven by authors noticing the 
    % (``if everyone is citing a publisher, it must be a legitimate publisher'').
    % This creeping increase in citations to potential predatory publishers could be called a silent predatory colonization of CHI.
Another explanation are the rising expectations of peer reviewers.
The submission guide at CHI 2016 explicitly mentioned that \textit{``lack of references to prior work is a frequent cause for complaint -- and low rating -- by reviewers''} \cite{2016guide}.
It is not uncommon for peer reviewers to demand inclusion of a long list of references in their reviews.
Authors often interpret these references not as suggestions but as mandatory for paper acceptance, and subsequently incorporate them into their paper.
A simple strategy of ``the more, the better'' could be motivating authors in CHI to~-- preemptively~-- include more references in their articles to satisfy the growing expectations of peer reviewers.

Likely, though, it is a % culmination
combination of several different co-occurring factors such as the ones presented in this article (i.e.,  the increase in systematic reviews in HCI, growing diversity of topics and co-authorship, growing expectations among peer reviewers, growing number of publications, and pressure to publish) that contributes to the increase in the mean number of references in CHI articles.
We leave a detailed qualitative investigation of the motivation of authors to include more references in their papers to future work (for instance with interview studies, workshops, or a Delphi study).
In the following section, we discuss why the growing number of references in articles is a critical concern that demands attention.%
\subsection{Implications % of the Observed Trends
for Authors, Reviewers, and the Broader Academic Community}%
References in scholarly articles serve an important function: they help novices in the field to
% An article with many references can help novices to
% and those unfamiliar with a research area
learn and become familiar with the research area.
However, we argue the observed trend (i.e., a culture of excessive citation, as evident in the rising number of references in CHI articles) is a negative development for authors, peer reviewers, the CHI community, and the entire field of HCI, for a number of different reasons.

% First, we argue that 130~references per article is too high a number to thoroughly screen during the peer review process.
% A high amount of references increases cognitive load during peer review and contributes to peer reviewer fatigue.
% This only adds to the already high amount of peer review fatigue, with conferences struggling to recruit high-quality reviewers.
% Peer review is the backbone of academic rigor.
% If reviewers throw only a fleeting glance at the reference section during peer review, then peer review is failing its purpose.
% % Peer reviewers will at best give the reference section a fleeting glance. This defeats the purpose.
% % Of course, these days one can already be happy if the peer reviewer reads the paper at all. But that is another story.
First, we argue 130~references per article are too many references to thoroughly screen during the peer review process. A high number of references increases the cognitive load on peer reviewers, contributing to reviewer fatigue. This only adds to the already high amount of peer review fatigue, with conferences struggling to recruit high-quality reviewers.
Peer review is the backbone of academic rigor, and if reviewers only give a cursory glance at the reference section, the integrity of the peer review process is compromised.%

% Second, the number of references included in CHI articles is a signal to the CHI community members. Peer pressure and herd mentality mandates that if the average is at a certain level, then an average article must include that amount of references. Deviations from the norm will likely be regarded as negative by peer reviewers.
Second, the number of references included in CHI articles serves as a signal to the CHI community. Peer pressure and herd mentality suggest that if the average number of references is high, authors feel compelled to include a similar number of references in their articles. Deviations from this norm are likely to be % viewed negatively
penalized by peer reviewers, creating a spiraling pattern of year-by-year increasing academic pressure and competition.
This spiraling pattern contributes to both author and peer reviewer dissatisfaction and fatigue.

% Third, even though the best paper awards and honorable mentions have moved away from awarding high-reference articles, the trend of the overall growing number of references in CHI articles still sends the wrong signals to junior members of the CHI community. Today, junior members are faced with a seemingly insurmountable amount of literature work. This distracts junior members from learning their craft and xxxxxxxxxxx.
Third, despite the shift away from awarding best paper awards and honorable mentions to high-reference articles,
% \ref{(cf. Section~\ref{sec:awards})},
the overall rise in the mean number of references per article still sends the wrong message to junior members of the CHI community.
The literature work required to publish an article at CHI poses a significant and undue burden on young scholars.
Today, PhD students in HCI are faced with a seemingly insurmountable and overwhelming amount of literature,
% to be incorporated into CHI papers,
which distracts the students from % focusing on
developing their research skills and expertise.

% Fourth, the inflation of citations in CHI articles acts as a signal to the HCI community.
% A citation today is worth less than it was in the past.
%     As an analogy, a citation is like a link from one web page to another. In the early days of search engine optimization, web developers tried to optimize the ``link juice'' flowing to their website, referring to the number of back links that their web page received from Google Search.
% The increasing number of citations xxxxxxxxxxxxxxxxx
Fourth, excessive citation contributes to inflation of the value of each citation.
Some authors may use ``citation stuffing'' to prop up the perceived value of their articles.
% Citation value is subject to inflation, since 
Due to this carelessness, a citation today does not carry the same weight it did in the past. This is similar to how early search engine optimization efforts devalued the significance of web links.
The increasing number of citations reduces their impact and contributes to
% can lead to 
% an overload
a hemorrhage
of less meaningful references.
% inflating important indicators, such as the ones listed on Google Scholar.

% Last, the high number of references makes it more easy for bad actors to commit fraud.
% In the presence of hypercompetition and ``perverse'' incentives \cite{edwards-roy-2017-academic-research-in-the-21st-century-maintaining-scientific-integrity-in-a-climate-of-perverse.pdf}, academic fraud is surprisingly wide-spread \cite{tijdink2014.pdf}.
% Without trade-off cost between citations in an article, the references section can be expanded to an arbitrarily high number of references.
% This facilitates fraudulent manipulation of citations and, at the same time, more difficult to detect.
% Citation manipulation can include self-citation \cite{11192_2010_Article_306.pdf}, author collusion in citation cartels \cite{s41598-021-93572-3.pdf,2005.14343.pdf}, and purchasing of citations \cite{2402.04607.pdf}.
Last, the high number of references makes it easier for bad actors to commit misconduct and fraud.
In an environment of hypercompetition and ``perverse'' incentives \cite{edwards-roy-2017-academic-research-in-the-21st-century-maintaining-scientific-integrity-in-a-climate-of-perverse.pdf}, academic fraud is surprisingly common \cite{tijdink2014.pdf}. % Oransky
With no clear trade-off cost between citations in an article, the reference section can be expanded arbitrarily, facilitating citation manipulation. This can include self-citation \cite{11192_2010_Article_306.pdf,TechReportV2.pdf}, author collusion in citation cartels \cite{s41598-021-93572-3.pdf,2005.14343.pdf},
    % review collusion/cartels: 3429776.pdf
and even the purchasing of citations \cite{2402.04607.pdf,else2021.pdf}.
The high number of references that, likely, undergo only cursory peer review makes it harder to detect such % academic misconduct
unethical and fraudulent research practices.%

% \todo{%
% This is particular important for peer review which lies at the foundation of academic credibility and which is becoming increasingly under pressure in recent years.
% }%

% In this rigid system entangled in ancient traditions, researchers are expected to volunteer for peer review on top of their ongoing workload.
% It is not surprising that scholarly peer review has gotten into a crisis in recent years, with knowledgeable peer reviewers being increasingly hard to engage in peer reviews, and high-quality reviews becoming scarce.

% It is not necessary to cite over 100 works in a paper to bring a point across. And it is the peer reviewers' job to judge the novelty of a paper (provided they have the required subject matter expertise).

\subsection{Reflective Considerations for Future Research Practices}%
\label{sec:recommendations}%

Many of the developments in the CHI community are plan-driven.
    For instance, the number of articles accepted at the CHI Conference is deliberately expanded each year.
One should not forget that such top-down decisions and policy changes lead to behavioral changes in members of the research community.
    % Notably
    However, while the CHI Steering Committee initially started the CHI community on its expansive path with its 2016 policy, it is the collective decision of authors at CHI to, year-by-year, cite more and more articles.
    In this section, we urge the community to pause and reflect on its citation practices.

This work is meant as a zealous vehicle for the members of the CHI research community to open their eyes to the community's growing problems.
We presented trends in citation practices to foster dialogue --- not to prescribe mandates, but to encourage diverse interpretations and applications of these findings. 
We invite readers to engage with the data, reflecting on its implications for the future of HCI research.
Our work serves not as a directive, but as a mirror reflecting the current state of CHI's scholarly practices.
We acknowledge that works like ours can be misinterpreted. We would like to explicitly stress that this work is not intended to be a ``guide'' or ``recipe'' for writing successful CHI papers.
By offering a data-backed overview of growing pain points in the field of HCI, we instead aim to help scholars and practitioners navigate the complexities of the modern research landscape and drive positive change in the community.

However, a critical question arises. How much responsibility does each author in the CHI community bear for the observed changes in citation practices?
While scientific knowledge advances rapidly, the academic system remains relatively conservative and resistant to change.
This system, dominated by powerful publishers like ACM Press, often feels too vast for any one individual to influence significantly.
Therefore, we also call for reflective consideration on an individual level.
We urge CHI authors to be more deliberate with their citations, viewing them as quality signals and endorsements. %, rather than empty fluff to prop up papers.
Readers should consider how citation trends impact their research practices and the broader HCI community.
% Questions like ``How do these trends influence the accessibility and inclusivity of HCI research?'' or ``What role do I, as a researcher, play in diversifying scholarly sources?'' are crucial.

Through introspection, the community can shape future research in ways that are thoughtful %, inclusive,
and aligned with shared values.
The introspective approach emphasizes the agency of the scholarly community to enact change, respecting the diversity of perspectives within our field and acknowledging each researcher’s autonomy to interpret and integrate these insights according to their goals.
We should acknowledge that, while fundamental goals (such as publishing articles in high-quality venues) are shared by all of us, the way to achieve these goals varies on an individual level, as demonstrated, for instance, by the `slow science' movement \cite{443271e.pdf}. This movement calls for a more deliberate and cautious approach to publishing.
Moving forward, the CHI community should find ways to accommodate these diverse approaches in its citation practices.

\subsection{Potential Solutions}%
\label{sec:solution}%
% --------------------------------------
In this section, we 
% provide % an opinion on
discuss potential solutions addressing the observed upward trend in the mean number of references in CHI articles.
% One solution in response to the observed trend is to simply ignore it.
% However, this would eventually lead to a situation where we wake up one day and wonder how we got there. It is more constructive to pro-actively discuss how the issue of growing number of references can be addressed.
One potential response is to simply ignore the trend, though this approach risks future challenges.
Proactively addressing the issue of 
% an increasing number of references
excessive citations
is a more constructive approach.
This section discusses four potential solutions % to mitigate this problem.
% how the issue of growing number of references can be addressed.
for % managing
addressing the growing number of references.

\subsubsection{Solution 1}
% \subsubsection{Policy change}
A first solution would be to revert to the 2015  policy (or an updated version thereof).
% This could be in the form of a limit on the number of references or a limit on the overall number of article pages (as was the case in the year 2015), or limits could be adapted to the current standards.
% Back in 2015, citing a paper was an endorsement. Any citation made to other works had to be weighed against the space it would take away from the article's text content or other references.
This could involve setting a limit on the number of references, the number of reference pages, or the overall number of pages per article, as was the case in 2015.
At that time, a citation was an endorsement, and any citation had to be weighed against the space it would take away from the article’s content or other references.
This trade-off has since been lost.
% This trade-off cost between citations has been lost. While it is more modern to allow an  unrestricted number of pages and references, in practice, there have been rather ill-defined and inconsistently enforced page limits imposed on CHI articles in recent iterations of the CHI Conference.
% The quality signaling by the Chairs during the CHI '23 submission process made clear that an unrestricted growth in article length is not welcomed by the Conference Chairs and the Steering Committee.
% Limiting the amount of references for most articles (with exception of literature reviews) would be a viable solution to address the growing number of references included in CHI articles.

While modern research practice tends to allow an unrestricted number of pages and references, recent iterations of the CHI Conference have seen strong nudging toward lower page limits.
For instance, recent CHI conferences clearly signaled that unrestricted growth in article length is not welcomed by the Conference Chairs and the Steering Committee.
The suggested number of words for CHI articles in 2024 was 7000--8000 words -- far lower than the average \cite{oppenlaender2023mapping}.
% During the CHI~’23 submission process % indicated
Yet, the strong recommendations are inconsistently enforced by the subcommittees.
Limiting the number of references for most articles (with the exception of % systematic
literature reviews) could be a viable solution to address the growing and excessive number of references included in CHI articles.
For such an approach to be successful, hard limits would need to be set, without exceptions, and these limits would need to be consistently enforced by all subcommittees.%

\subsubsection{Solution 2}
A second solution is to develop language models for supporting paper writing.
In particular, specialized language models could be trained to automatically cite prior work, given the text of a manuscript.
From the perspective of the author, this would alleviate much of the literature work, and it would help authors concentrate on ideation and conducting high-quality work.
However, as the failed release of Meta's Galactica model demonstrated in 2022 \cite{galactica}, language models are still prone to hallucinate non-factual and biased information which makes these models, in their present form, not suitable for literature work.
\transition{%
This solution would also not alleviate peer reviewer fatigue, which we discuss in the next section.}%

\subsubsection{Solution 3}
% \subsubsection{Socio-technical solution}
A third solution pertains to the nature of the peer review process in academia.
Peer review is the cornerstone of scholarly rigor, serving as a crucial filter that ensures the quality, relevance, and integrity of academic publications.
From the perspective of creativity theory~\cite{2012RuncoJaegerStandardDefinition.pdf} and Csikszentmihalyi's systems model of creativity~\cite{Csikszentmihalyi},  the peer review process is a gatekeeping process that validates two critical components of an article: novelty and importance (also referred to as usefulness or appropriateness) \cite{2012RuncoJaegerStandardDefinition.pdf}.
    % The task of the peer reviewer is to validate the novelty and usefulness of the article by comparing the article's contribution to existing works in the field.
    % , and to assess the usefulness of the methodologies, findings, and contributions of research before dissemination to the broader community.
    % First, it assesses an article's novelty. The task of the peer reviewer is to validate the novelty of the article by comparing the article's contribution to existing works in the field.
    % Second, peer review assesses the usefulness (or appropriateness) of the methodologies, findings, and contributions of research before dissemination to the broader community.
    % Novelty and usefulness go hand in hand, as a contribution may be novel, but irrelevant or useless for the field.
As gatekeeper, it is the peer reviewer's task to assess these two components and, hence, the creativity of a submitted article.
% Both of of the two above components of creativity need to be evaluated for effective peer review.
%
% In the dynamic and interdisciplinary field of HCI, peer review is especially vital. Peer review fosters a culture of scrutiny and debate, propelling the field forward by ensuring that only work that meets a high standard of scholarly excellence is published. This process reinforces the credibility of individual research efforts and maintains the overall trustworthiness of the academic corpus.
% Peer review, by demanding a critical evaluation from independent experts, underscores the collaborative nature of science, where knowledge is continually built upon, refined, and expanded.
%
In some cases, an extensive related work section may support the evaluation of an article's novelty, especially in cases where the peer reviewer may not have sufficient knowledge for conducting the review.
    But in an ideal world, a peer reviewer lacking sufficient knowledge of the field should not review an article in the first place.
    In practice, however, a mismatch between reviewer expertise and a paper is, unfortunately, rather common \cite{3528086.pdf}, especially in large and highly diverse conferences, such as CHI.
% However,
In the role of a gatekeeper, it is the peer reviewer's duty to be knowledgeable about works in the field.
% However,
In a field as vast and diverse as HCI, this can be challenging.

A potential solution % to addressing the above issue
is to split peer review into two parts, one conducted by the human peer reviewer, the other one performed by a machine.
%
% Note that we do not argue for fully automating the peer review process. However, large language models could be developed to automate the evaluation of novelty. Language models already exhibit above human-levels of performance in some areas. Because language models are knowledge compressors, they could perform the evaluation of an article's novelty and the completeness of the reference section much better than human reviewers.
Language models are knowledge compressors and can assess whether a contribution is original and whether the references are comprehensive more effectively than human reviewers.
This would allow human reviewers to focus on assessing the second component of creativity: the importance % usefulness and appropriateness
of the methodologies, findings, and contributions for the field.
    % Specialized language models could, with ease and in record time, assess whether a contribution has been made before, whether an idea has been articulated before, or whether an article accurately acknowledges prior work.
% This would leave the second part of the evaluation (usefulness or appropriateness) to the human reviewer.
    % The usefulness of an article's contribution is likely to be readily apparent even to non-expert reviewers, requiring less knowledge of the works in the field.
    % Besides usefulness, this part of the evaluation also would include non-functional requirements and other factors, such as aesthetics.

% The writing for this to happen is already on the wall.
% While it is easy to dismiss this proposed solution as science fiction, several tell-tale signs point toward the fact that we are heading in this direction.
Evidence suggests we are already moving toward this solution. % direction.
In the United States, automated essay scoring systems are already grading student essays~\cite{Texas} % Essays.pdf
and a growing number of peer reviewers use language models to write their reviews.
    A study by \citeauthor{2403.07183.pdf} found that up to 17\% of peer reviews at top machine learning conferences are substantially authored by language models~\cite{2403.07183.pdf}.
    \citeauthor{2310.01783.pdf}'s study found an overlap of between 30.9\% and 39.2\% between the points raised by GPT-4 and human reviewers, and their user study with 308 researchers from 110 institutions in the US highlighted that researchers find AI-generated feedback useful.
% , and a large percentage (82.4\%) found the LLM-generated feedback more useful than feedback from some human reviewers.
    % with many researchers finding AI-generated feedback more useful than some human feedback \cite{2310.01783.pdf}. 
    Similarly, \citeauthor{3637371.pdf} found that automated peer review significantly expedited the review process and improved review coverage, although raising concerns around trust, bias, and agency \cite{3637371.pdf}.

Automating aspects of peer review could offer significant benefits \cite{2405.06563.pdf}.
The cost of peer review is estimated to be billions of dollars annually~\cite{s41073-021-00118-2.pdf}. In a system strained by an overwhelming volume of literature \cite{2309.15884.pdf}, partial automation can alleviate the burden on reviewers.
Further, automating peer review could increase trust in the scientific process by reducing biases, such as the Matthew effect, which has been a concern in academia since at least 1968 \cite{merton1968.pdf}.
In the future, language models could assist in evaluating novelty, leaving %  the assessment of usefulness
other aspects to human reviewers,
thereby enhancing the efficiency and reliability of the peer review process.%
\subsubsection{Solution 4}%
The potential solutions discussed so far have centered around policies
% mandates
and peer review.
Another idea focuses on driving change from the author side.
In some disciplines, journals require authors to pre-register their studies before conducting them.
What if CHI authors % at the  Conference
pledged to submit their articles with a % fixed
pre-determined number of references?
Authors could set this limit for themselves and they would pre-register this limit publicly before writing and submitting their paper.
The pre-registered limit would work as a self-binding contract, forcing authors to keep within their % promised
pledged number of references.
This -- like in the years before the page limit was lifted -- would force authors to make a trade-off between references and to think more carefully how many and which references to include in their articles.
To incentivize participation in this pre-registration scheme, awards could be given to authors who conform to their pre-registered number of references.
The public dataset of pre-registrations could % implicitly
contribute to establishing % agreeing on
new community norms and standards for the number of references in CHI articles.
However, these standards would not need to be the same for every author. For instance, students could be allowed to include fewer references, acknowledging their role as junior members in the research community. This would alleviate the undue burden of literature work on young scholars.%
%
%
%
%
%
%
%
%
% ===========================
\section{Conclusion}%
\label{sec:Conclusion}%
% ===========================
% Summarize the main findings and their significance.
% Reiterate the call for action or change in practices, integrating the forward-looking perspective presented.
This article presented meta-research on citation practices at the ACM CHI Conference, the top conference in the field of Human-Computer Interaction.
% \todo{While not a complete substitute, CHI serves as a proxy for broader developments in the field of HCI in this work.}
Our % segmented regression
exploratory analysis provided clear evidence of a change of pace in academic working in the CHI community, indicative of broader trends in HCI.
We provided empirical evidence of a year-by-year increase in the number of references included in CHI articles.
The % field's
excessive citation practices were enabled by a policy decision that lifted page restrictions on the reference section of CHI articles.
This policy decision constitutes a non-linearity in the CHI community's trajectory and had a destabilizing effect on the  community.
% In this paper, we shed light on the effect of a change in policy and several factors that potentially contribute to the observed trend.
% Both authors and reviewers are overwhelmed by the observed trends, with negative consequences for the HCI community.
The implications of this meso-level policy decision are profound, 
% This work highlights
underlining the value of meta-research for the CHI community and HCI.
% with qualitative and quantitative behavioral changes.
More meta-research is needed to carefully consider the broader impact of policy decisions and to provide clear recommendations for stakeholders. %~\cite{meta-HCI}.
%
% We urge the community to pause and reflect on these trends.
It is time for the HCI community to pause and reflect on its citation practices and constructively discuss solutions.
% Whether it is the solution proposals made in this article or other solutions,
A different pathway forward is urgently needed.%
% to make the HCI field sustainable again.
% , in particular with an eye on the research community's junior members and the sustainability of the HCI field.
% Let's also include our future selves in the discussion and today's junior members of the HCI community who will be the future leaders, but only if sustainable pathways forward are provided.
%
%
% Often, for change to take place, one needs to radically break with the past.
% % The iPhone revolutionized the mobile phone industry. The Tesla cybertruck revised what is possible in the design of production-line vehicles.
% In academia, in particular, tradition and customs dictate how academic work is being conducted today.
% This stands in contrast with the requirement of many funding organizations demanding justifications on how (planned) research contributes to the ``renewal'' of science.
%     The term ``renewal of science'' in the context of research funding refers to initiatives, discoveries, or methodologies that significantly advance scientific knowledge, introduce new paradigms, challenge existing theories, or apply existing knowledge in novel and impactful ways.
% This paper is a call to radically break with the past and the current trends of academia.
%
%
%
\begin{acks}
% \subsection*{Acknowledgements}%
\iftoggle{anonymous}{The author(s)}{The author}
aimed to limit this paper to 35~references, but unfortunately, this target was exceeded.
\iftoggle{anonymous}{We pledge}{I pledge}
 to % improve
do better
next time.%
\end{acks}
\balance%
\bibliographystyle{ACM-Reference-Format}
\bibliography{paper}
% 
% \printbibliography

% \appendix
% \section{Interview Guides}

\end{document}

%% file: FIG-REFS-FULL.tex
%TC:ignore
% \begin{figure}[!htb]
%   \centering
%   \includegraphics[width=\linewidth]{figures/references-2015.pdf}
%   \caption{Number of references per CHI article from 2015 to 2023.}
%   \Description{Number of references per CHI article from 2015 to 2023.}
%   \label{fig:boxplots}
% \end{figure}
\begin{figure}[!bht]%
  \centering%
  \includegraphics[width=\linewidth]{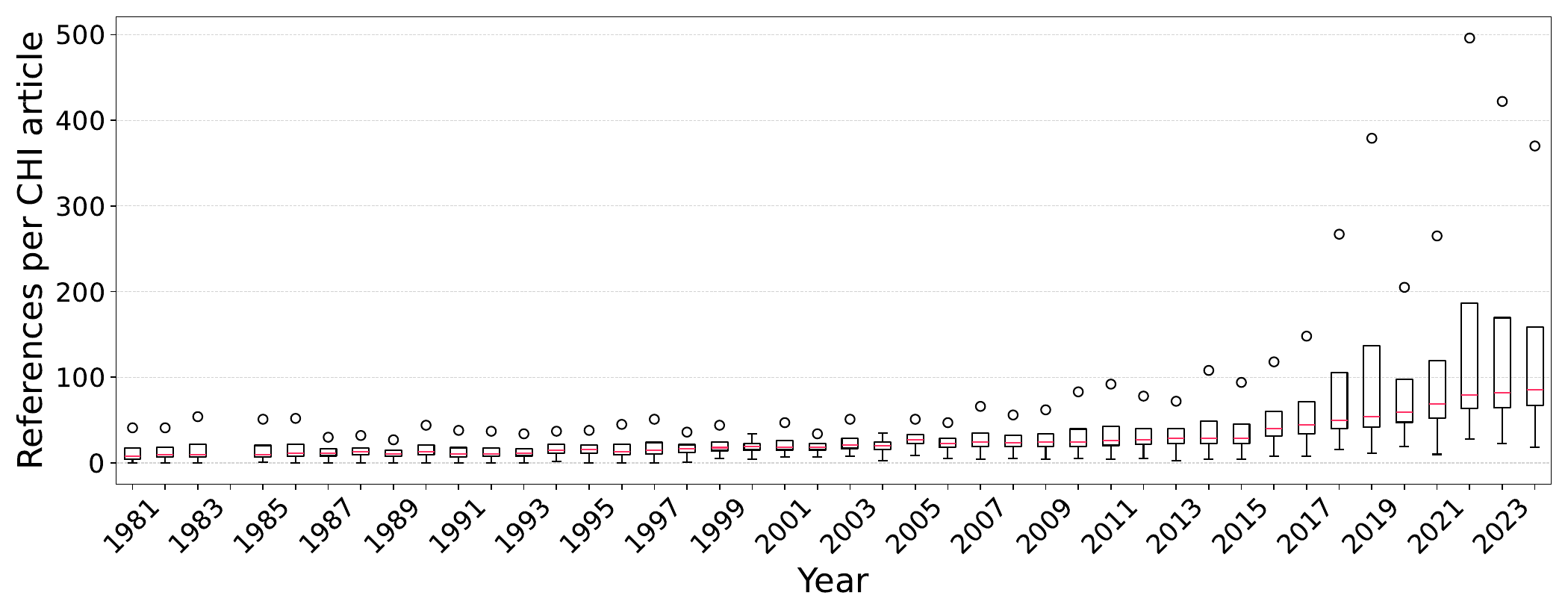}%
  \caption{Number of references per article in the CHI Conference Proceedings from 1981 to 2024.}%
  \label{fig:boxplots}%
\end{figure}%
%TC:endignore

%% file: FIG-CFPs.tex
%TC:ignore
\begin{figure*}[!thb]%
\centering%
  \begin{minipage}{0.47\textwidth}
    \centering
    2015 CfP (excerpt)\\
    \fbox{ % This creates a framed box
      \begin{minipage}{\linewidth - 2\fboxsep - 2\fboxrule}
      % \small
      \textsf{%
        A Paper is no more than 10 pages long, while a Note is no more than 4 pages long. \textbf{This includes} figures, \textbf{references}, appendices and an abstract of less than 150 words long. Over length submissions will be rejected.
        }%
        \\
      \end{minipage}
    }
% \\[.5em]
  \end{minipage}%
  \hspace{.5cm}%
  \begin{minipage}{0.47\textwidth}
    \centering
    2016 CfP (excerpt)\\
    \fbox{ % This creates a framed box
      \begin{minipage}{\linewidth - 2\fboxsep - 2\fboxrule}
      % \small
      \textsf{%
        A Paper is no more than 10 pages long, while a Note is no more than 4 pages long. \textbf{References do not count toward these lengths.} The lengths do include figures, appendices, and an abstract of less than 150 words. Submissions that exceed these limits will be rejected.
        }%
      \end{minipage}
    }
  \end{minipage}%
  % \\
  % \begin{minipage}{0.47\textwidth}
  %   2015 Submission Guide (excerpt)\\
  %   \fbox{ % This creates a framed box
  %     \begin{minipage}{\linewidth - 2\fboxsep - 2\fboxrule}
  %     \small
  %       (n/a)\\
  %       \vspace{4em}
  %     \end{minipage}
  %   }
  % \end{minipage}%
  % \hspace{.5cm}%
  % \begin{minipage}{0.47\textwidth}
  %   2016 Submission Guide (excerpts)\\
  %   \fbox{ % This creates a framed box
  %     \begin{minipage}{\linewidth - 2\fboxsep - 2\fboxrule}
  %     \small
  %         ``Relevant previous work: is prior work adequately reviewed?''\\
  %         \ldots\\
  %         \ldots ``letting others build on your work is the entire purpose of a CHI Paper or Note.''
  %     \end{minipage}
  %   }
  % \end{minipage}%
  \caption{\rev{Excerpts} from the Call for Papers (CfPs) at the CHI Conference in 2015 and 2016 (own highlighting).}%
  \label{fig:cfp-excerpts}%
\end{figure*}%
%TC:endignore

%% file: TAB-REGRESSION.tex
%TC:ignore
\begin{table}[!htbp]%
\caption{%
% Simple regression models fitted before and after the page restrictions were lifted at the CHI Conference.
% Models M1 and M2 are multiple regression models predicting the mean number of references in articles ($y$)
% in the presence of \todo{five} % confounding
% variables
% (see Section \ref{sec:context}):
% \todo{%
%     number of authors per paper ($c_1t$),
%     number of citations to arxiv ($c_2t$),
%     number of citations to data and code repositories ($c_3t$),
%     number of literature reviews ($c_4t$),
%     and
%     number of citations to questionable publishers
%     in CHI articles ($c_5t$).
% }%
Simple linear regression models
% (without confounding variables)
% to
evaluate the extent to which Year ($t$) can predict the mean (LM\textsubscript{1}) and minimum (LM\textsubscript{2}) number of References ($y$) in CHI articles.
The two models are used to predict the number of references per article in future CHI proceedings in \autoref{fig:predictions}.
}%
\label{tab:linear-regression-stats}%
% \small
\begin{tabularx}{\linewidth}{lXX}%
\toprule
     &  LM\textsubscript{1} \newline (Mean number of references)  &
      LM\textsubscript{2} \newline (Min number of references)
      \\
\midrule
    Year range & 2016--2024 & 2016--2024 \\
    Regression equation &
    \small
    $y={\beta_0} + {\beta_1}{\mathrm{t}}$ % + {\epsilon_t}$
    &
    \small
    $y={\beta_0} + {\beta_1}{\mathrm{t}}$ % + {\epsilon_t}$
    \\
    Coefficient $\beta_0$ (intercept) &
        -12971.92
        &
        -3620.33
    \\
    Coefficient $\beta_1$ (slope) &
        6.45
        &
        1.80
    \\
    t (for slope $\beta_1$) &
        $t=17.58$
        \newline
        $p<1e^{-6}$
        &
        $t=2.61$
        \newline
        $p=0.03$
    \\
    % df\textsubscript{regression}
    %     &
    %     1
    %     &
    %     1
    % \\
    % df\textsubscript{residual}
    %     &
    %     7
    %     &
    %     7
    % \\
    MAE &
        1.91
        &
        3.99
    \\
    MSE & 
        6.29
        &
        22.18
    \\
    R\textsuperscript{2} &
        0.98
        &
        0.49
    \\
    F-test &
        $F(1,7)=308.96$
        \newline 
        $p<1e^{-6}$
        &
        $F(1,7)=6.81$
        \newline 
        $p=0.03$
    \\
\bottomrule
\end{tabularx}%
\end{table}%
%TC:endignore

%% file: FIG-PREDICTION.tex
%TC:ignore
\begin{figure*}[!htbp]%
  \centering%
  \includegraphics[width=\linewidth]{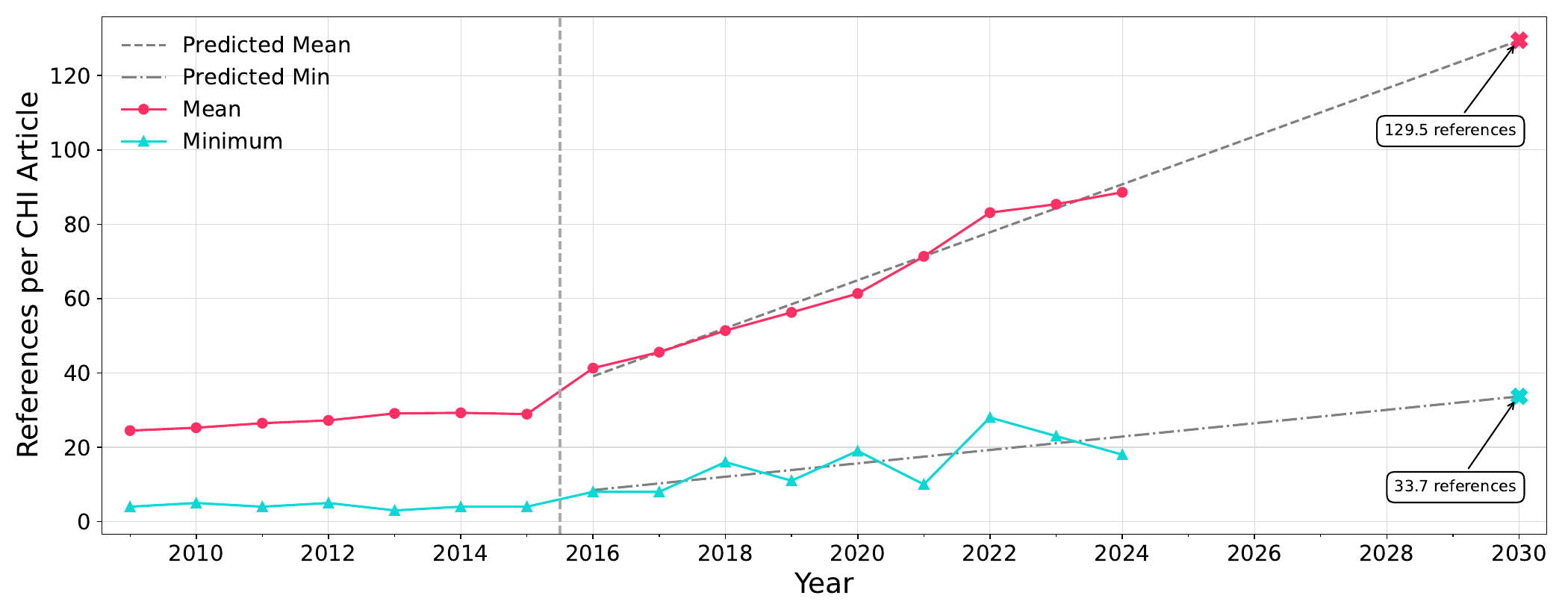}%
  \caption{Extrapolation of the observed trend in the mean (\textcolor{RED}{red}) and minimum (\textcolor{TEAL}{blue}) number of references per CHI article. CHI articles are
  % \rev{estimated}
  predicted
  to include on average almost 130 references in the year 2030. The minimum number of references per CHI~article is estimated to reach 33.7 references per article in 2030, which is more than the 2015 average level of references.}%
  \label{fig:predictions}%
\end{figure*}%
%TC:endignore

%% file: TAB-AWARDS.tex
%TC:ignore
\begin{table}[!htbp]%
\caption{Comparison of mean number of references in articles awarded a best paper award or honorable mention (M\textsubscript{awarded}) with the mean number of references in unawarded articles (M\textsubscript{regular}) in the CHI Proceedings from 2007 to 2024.}%
\label{tab:awards}%
\small%
\centering
\begin{tabular}{rrrrrrr}%
\toprule%
     Year &
     $M\textsubscript{awarded}$ &
     $M\textsubscript{regular}$ &
     $t$ & $CI$ (95\%) & $d$ & $p$  \\
\midrule
%%% If a p-value is less than 0.05, it is flagged with one star (*). If a p-value is less than 0.01, it is flagged with 2 stars (**). If a p-value is less than 0.001, it is flagged with three stars (***).
    2007   & 29.966  & 23.368 &  2.860 & [2.045, 11.149] & 0.580  & 0.005  * \\ % **
    2008   & 28.029  & 22.598 & 2.632  & [1.364, 9.499] & 0.491  & 0.009  * \\ % **
    2009   & 28.106  & 23.739 & 2.453  & [0.863, 7.872] & 0.393  & 0.015  * \\ % *
    2010   &  29.138 & 24.181 &  2.999 & [1.704, 8.21] & 0.420  & 0.003  * \\ % **
    2011   & 25.972  & 26.582 & -0.369  & [-3.859, 2.64] & -0.048  & 0.713   \hphantom{*} \\
    2012   &  27.923 & 27.069 & 0.496  & [-2.534, 4.242] & 0.068  &  0.620  \hphantom{*} \\
    2013   &  30.213 & 28.772  & 0.978  & [-1.457, 4.339] &  0.118 & 0.329  \hphantom{*}  \\
    2014   & 31.398  & 28.66 &  2.036 & [0.095, 5.381] &  0.227 & 0.042  * \\ % *
    2015   & 31.219  & 28.153 & 2.298  & [0.445, 5.685] & 0.243 & 0.022 * \\ % *
\multicolumn{7}{c}{%
    \noindent
    \begin{tikzpicture}    
        % Draw the line
        \draw[line width=0.5pt] (0,0) -- (.6\linewidth,0);
        % Place the word "Event" in the center with a white background
        \node[fill=white, inner sep=2pt] at (.6\linewidth/2,0) {Policy Change};
    \end{tikzpicture}
}%
    \\
    2016   & 43.470  & 40.705 & 1.616  & [-0.597, 6.127]
    & 0.170
    & 0.107   \hphantom{*} \\
    2017   & 49.884  & 44.497 & 2.779  & [1.581, 9.194]
    & 0.283
    & 0.006 * \\ % **
    2018   & 57.865  & 49.880 & 3.629  & [3.665, 12.307]
    & 0.369
    & 0.000 * \\ % ***
    2019   & 61.142  & 54.991 &  2.616 & [1.534, 10.768]
    & 0.242
    & 0.009 * \\ % **
    2020   & 66.051  & 60.125 &  2.678 & [1.582, 10.271]
    & 0.241
    & 0.008 * \\ % **
    2021   & 72.965  & 70.987 &  0.691 & [-3.642, 7.598]
    & 0.064
    & 0.490  \hphantom{*} \\
    2022   & 87.922  & 81.933 &  1.520 & [-1.747, 13.725]
    & 0.150
    & 0.129  \hphantom{*} \\
    2023   & 88.969 & 84.774 & 1.222   & [-2.544, 10.934]
    & 0.117
    & 0.222  \hphantom{*} \\
    2024   & 92.958 & 87.651 & 1.795   & [-0.494, 11.109]
    & 0.144
    & 0.073  \hphantom{*} \\
\bottomrule%
\end{tabular}%
\end{table}%
%TC:endignore

%% file: FIG-AUTHORS1.tex
%TC:ignore
\begin{figure*}[!thbp]%
  \centering%
  \includegraphics[width=\linewidth]{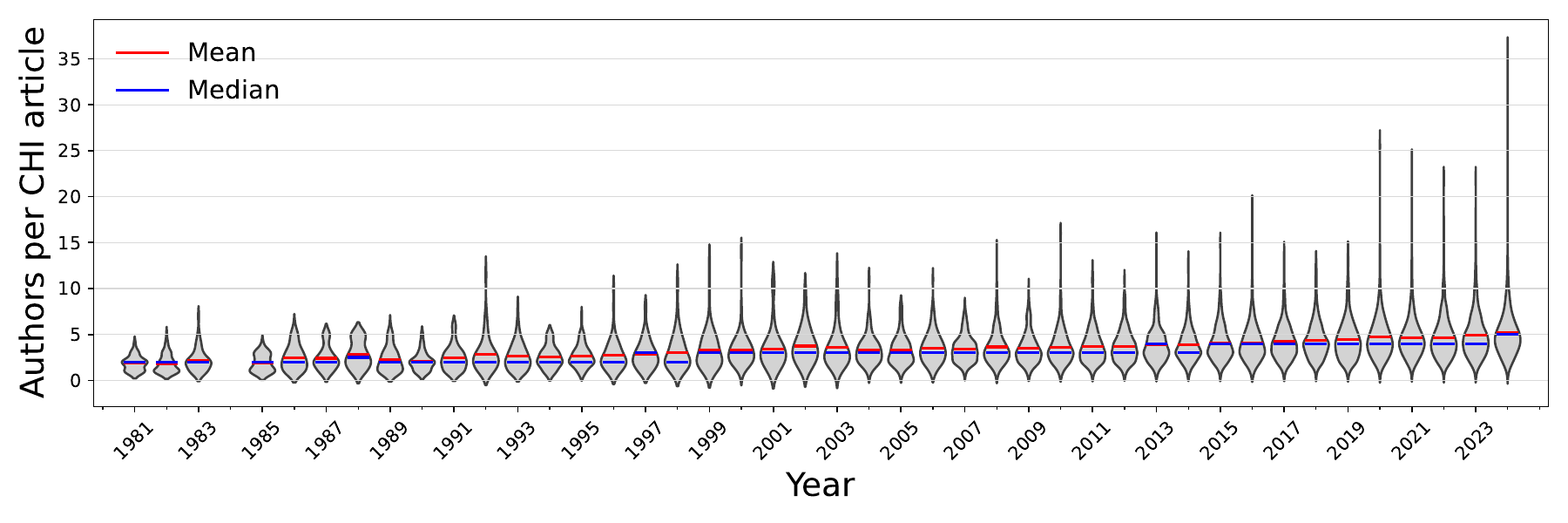}%
  \caption{\rev{The number of co-authors in CHI papers has increased over time, highlighting the trend toward increased collaboration in HCI.}}%
  \label{fig:authors-growth}%
\end{figure*}%
%TC:endignore

%% file: FIG-AUTHORS2.tex
%TC:ignore
\begin{figure*}[!htbp]%
\centering%
% \begin{subfigure}[b]{0.49\textwidth}
  \begin{minipage}{0.49\textwidth}
         \centering
  \includegraphics[width=\linewidth]{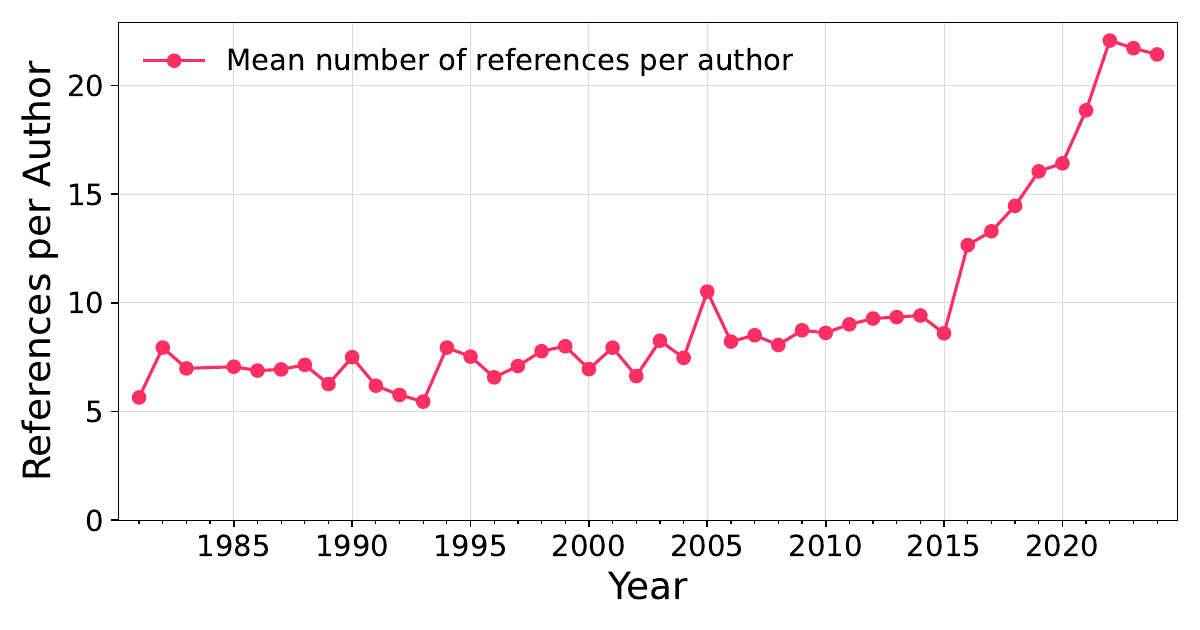}%
  \caption{\ok{Mean number of references per author in CHI articles\\~}}%
  \label{fig:refs-per-author}%
% \end{subfigure}
% \hfill
% \begin{subfigure}[b]{0.49\textwidth}
\end{minipage}
\hfill
\begin{minipage}{0.49\textwidth}
         \centering
  \includegraphics[width=\linewidth]{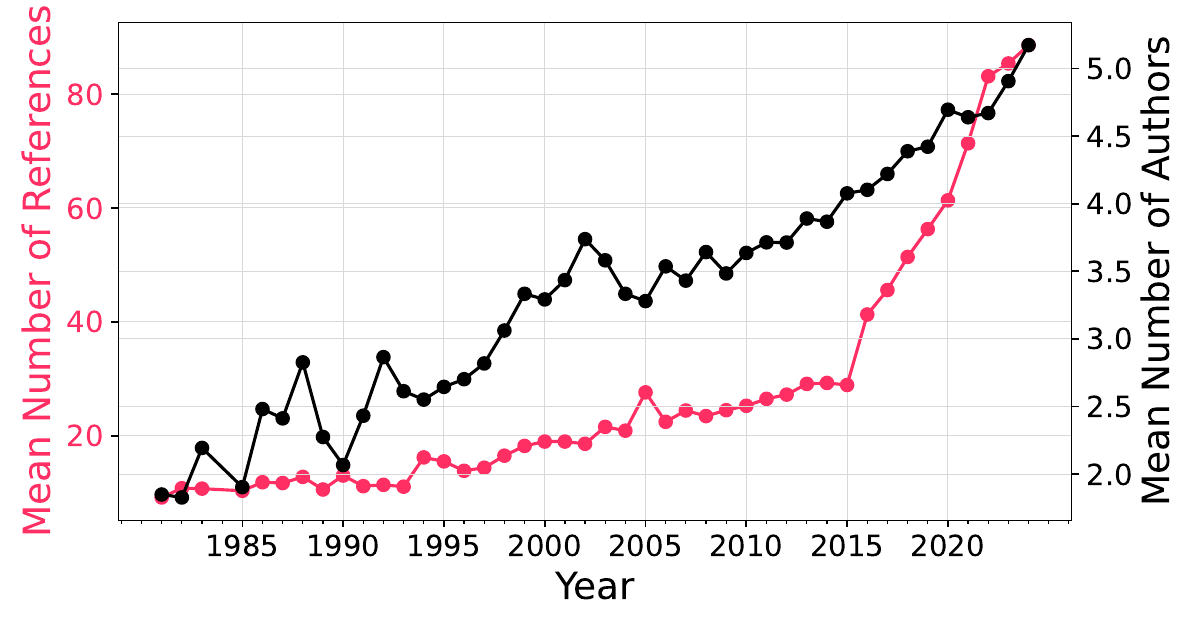}%
  \caption{\rev{Mean number of references (primary axis) and mean number of authors (secondary axis) per CHI article.}}%
  \label{fig:authorcitationcomparison}%
% \end{subfigure}
%   \caption{\rev{TODO.}}%
%   \Description{TODO.}%
%   \label{fig:authors-growth}%
\end{minipage}
\end{figure*}%
%TC:endignore

%% file: FIG-LITREVIEWS.tex
%TC:ignore
\begin{figure*}[!thbp]%
  \centering%
% \begin{subfigure}[b]{0.49\textwidth}
  \begin{minipage}{0.49\textwidth}
         \centering
  \includegraphics[width=\linewidth]{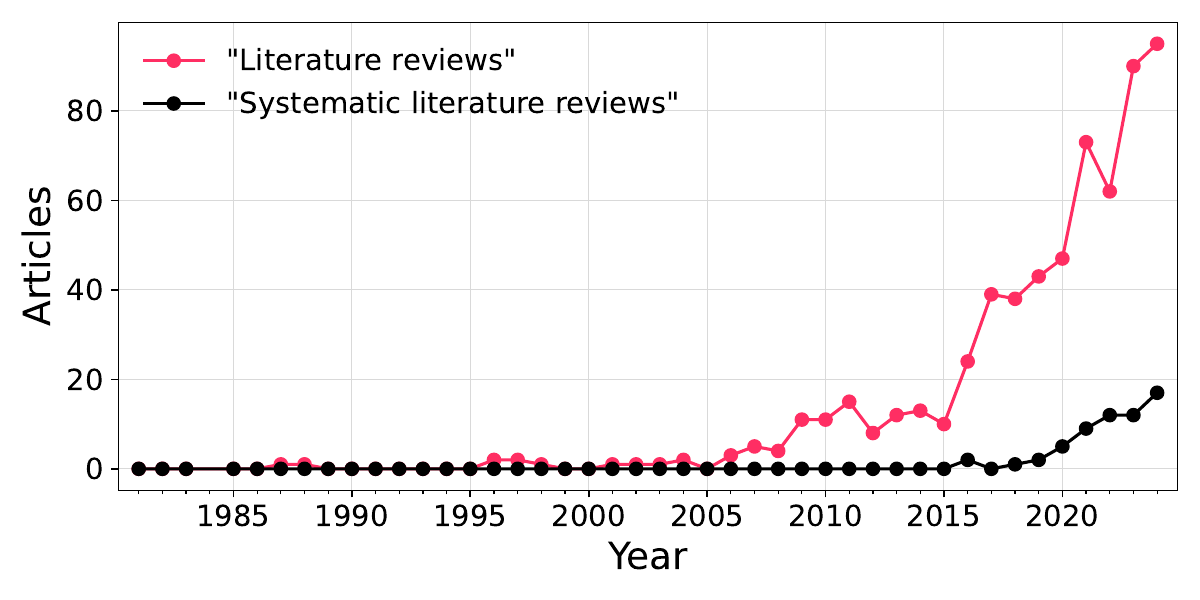}%
  \caption{\rev{Absolute number of articles in the CHI proceedings reporting having conducted a literature review}}
  \label{fig:litreviews:a}%
% \end{subfigure}
% \hfill
% \begin{subfigure}[b]{0.49\textwidth}
  \end{minipage}
  \hfill
  \begin{minipage}{0.49\textwidth}
        \centering
  \includegraphics[width=\linewidth]{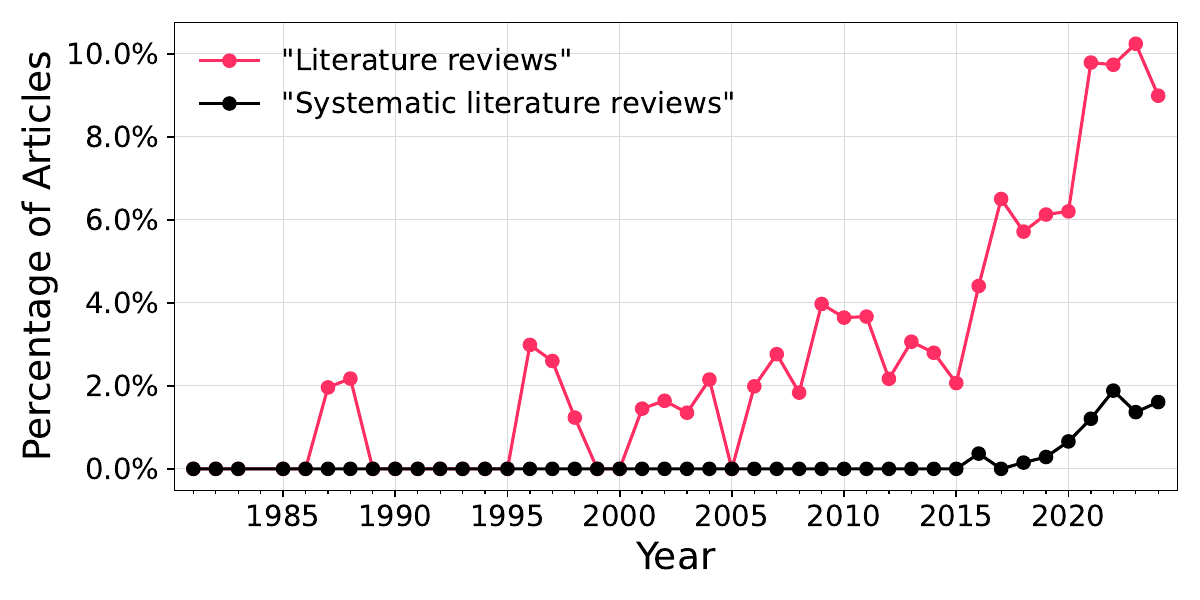}%
  \caption{\rev{Relative number of articles in the CHI proceedings reporting having conducted a literature review}}
  \label{fig:litreviews:b}%
% \end{subfigure}
\end{minipage}
%   \caption{Absolute \rev{(a)} and relative \rev{(b)}
%   \rev{number of articles in the CHI proceedings reporting % having conducted
%   a ``literature review'' (including ``literature surveys'' and ``systematic literature reviews''; in red) and a ``systematic literature review'' (in teal color) has increased since 2015.} Systematic literature reviews in CHI articles are a new phenomenon \rev{at CHI.}
%   % starting in 2016.
%   }%
%   \Description{Line plots of literature reviews in CHI proceedings from 1981 to 2024 show that literature reviews in the CHI proceedings have become more common since the editorial change at CHI '16, both in absolute and relative terms.}%
% \label{fig:litreviews}%
\end{figure*}%
%TC:endignore

%% file: FIG-ARXIV.tex
%TC:ignore
\begin{figure*}[thbp]%
\centering
% \begin{subfigure}[b]{0.49\textwidth}
\begin{minipage}{0.49\textwidth}
         \centering
  \includegraphics[width=\textwidth]{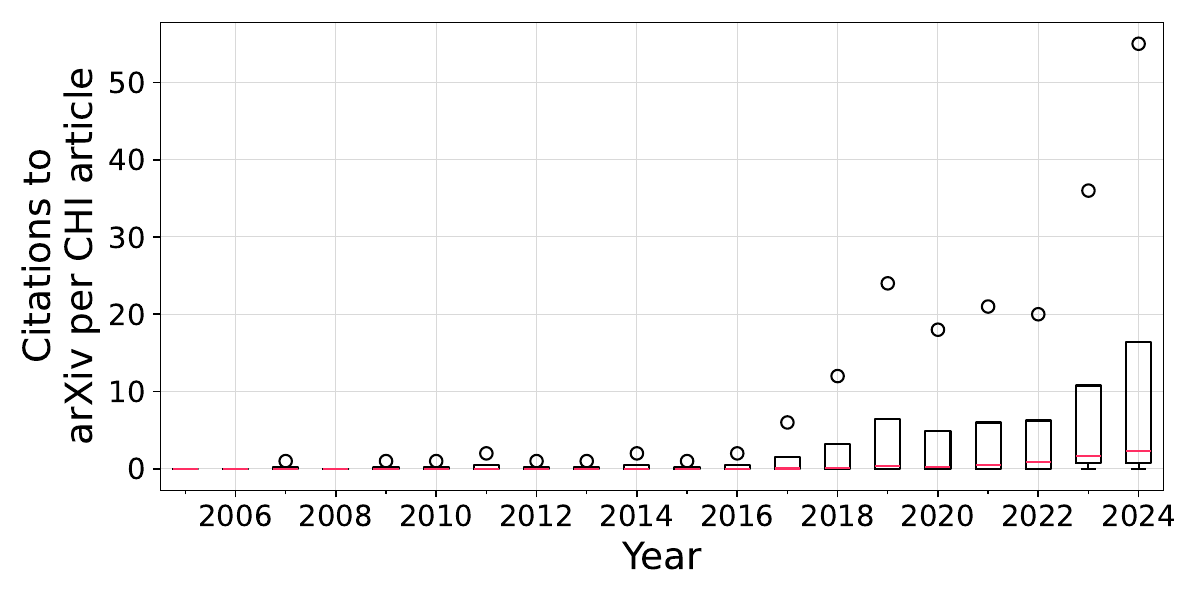}%
  \caption{Formal citations to arXiv pre-print papers in CHI articles}
  \label{fig:arxiv}%
\end{minipage}
\hfill
\begin{minipage}{0.49\textwidth}
         \centering
  \includegraphics[width=\textwidth]{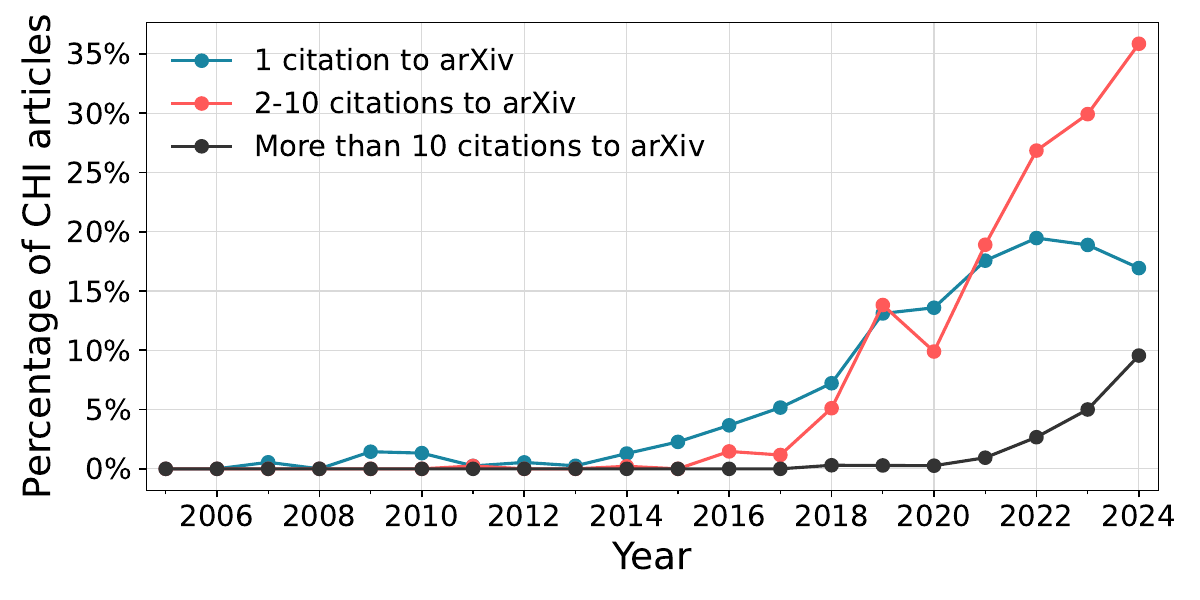}%
  \caption{Relative number of articles formally citing arXiv pre-prints $n$ times}
  \label{fig:arxivnum}%
% \end{subfigure}
  \end{minipage}
% \caption{Citations to pre-print arXiv papers (a) and percentage of CHI articles citing arXiv pre-prints $n$ times (b) in the CHI Proceedings. % from 1981 to 2024.
% }
% \Description{Citations to pre-print arXiv papers (a) and percentage of CHI articles citing arXiv pre-prints $n$ times (b) in the CHI Proceedings.}%
% \label{fig:three graphs}
\end{figure*}
%TC:endignore

%% file: FIG-REPOS.tex
% \begin{figure}[!htb]%
%   \centering%
% \begin{subfigure}[b]{0.49\textwidth}
\begin{figure*}[!thbp]
  \centering
  \begin{minipage}{0.49\textwidth}
        \centering
  \includegraphics[width=\textwidth]{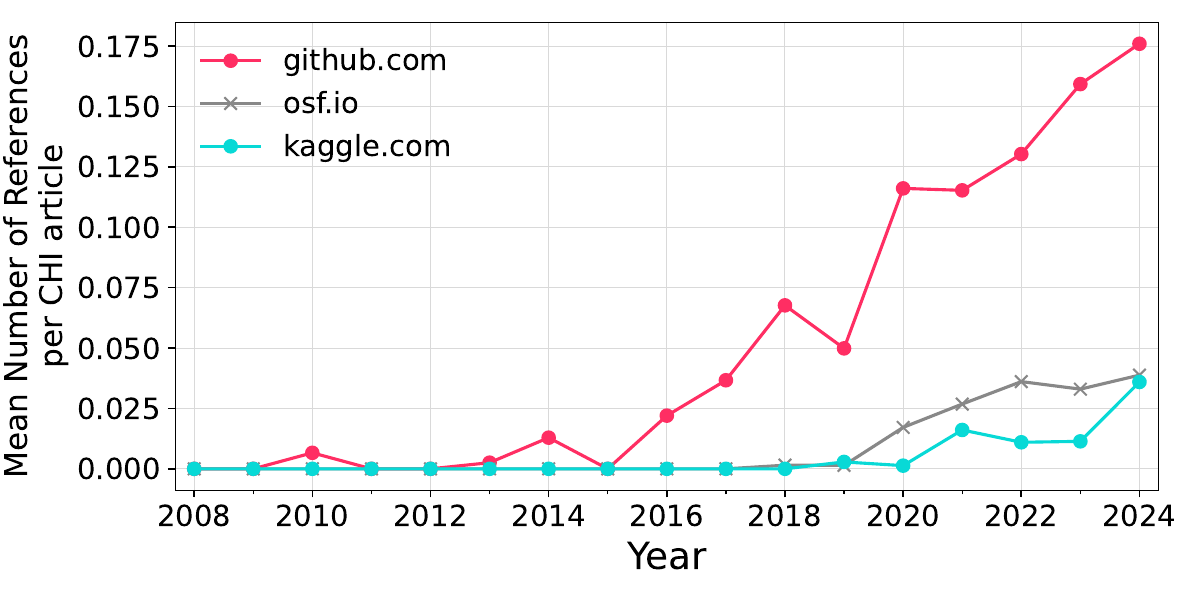}%
  \caption{\rev{Mean number of citations to data and code repositories per CHI article}}
  \label{fig:datarepos1}%
% \end{subfigure}
% \hfill
% \begin{subfigure}[b]{0.49\textwidth}
 \end{minipage}
  \hfill
  \begin{minipage}{0.49\textwidth}
         \centering
  \includegraphics[width=\textwidth]{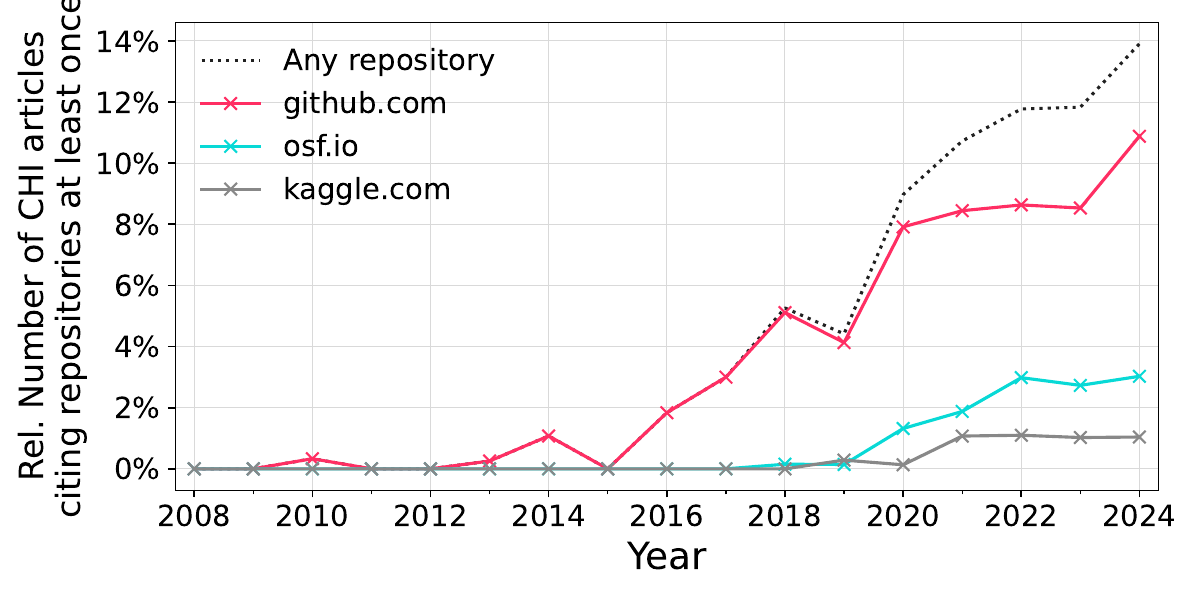}%
  \caption{\rev{Relative number of CHI articles citing at least one data or code repositories}}
  \label{fig:datarepos2}%
% \end{subfigure}
%   \caption{Mean number of citations to code and data repositories (a) and relative number of CHI articles citing at least one repository (b) in the CHI proceedings from 2008 to 2024.}%
%   \Description{Mean number of citations to data and code repositories (a) and relative number of CHI articles citing at least one repository (b) in the CHI proceedings from 2008 to 2024.}%
%   \label{fig:datarepos}%
  \end{minipage}%
\end{figure*}%

%% file: FIG-PREDATORS.tex
%TC:ignore
\begin{figure*}[!thbp]%
  \centering%
\begin{subfigure}[b]{0.32\textwidth}
     \centering
  \includegraphics[width=\textwidth]{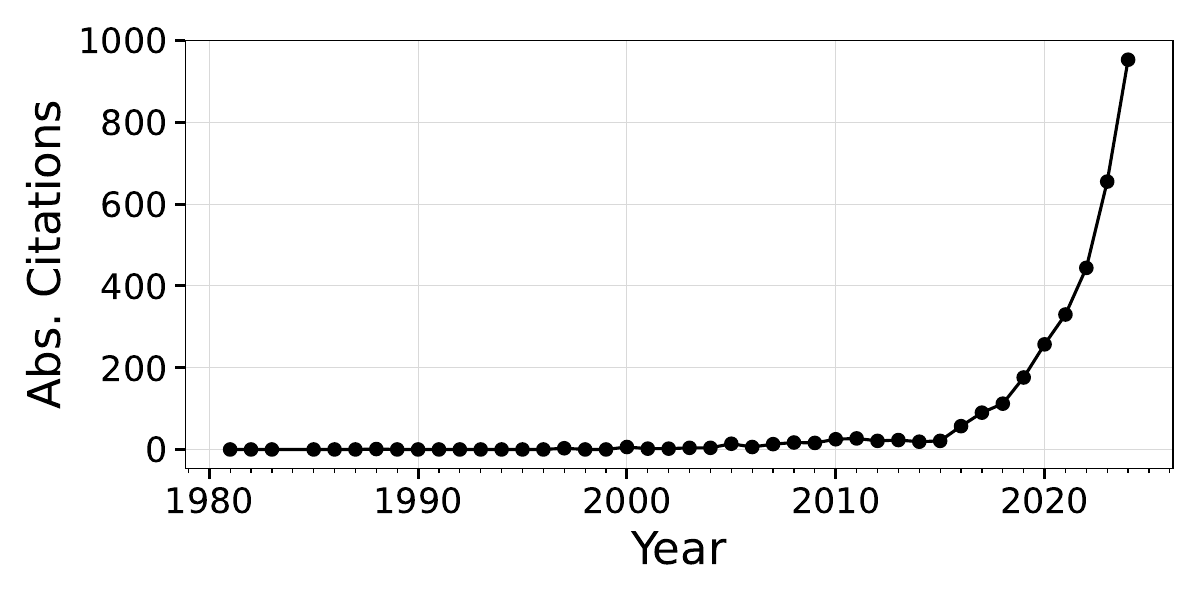}%
\captionsetup{labelformat=empty}
  \caption{(a) Absolute number of citations to potential predatory journals and publishers\\}
  \label{fig:beall-a}%
\end{subfigure}
\hfill
\begin{subfigure}[b]{0.32\textwidth}
     \centering
  \includegraphics[width=\textwidth]{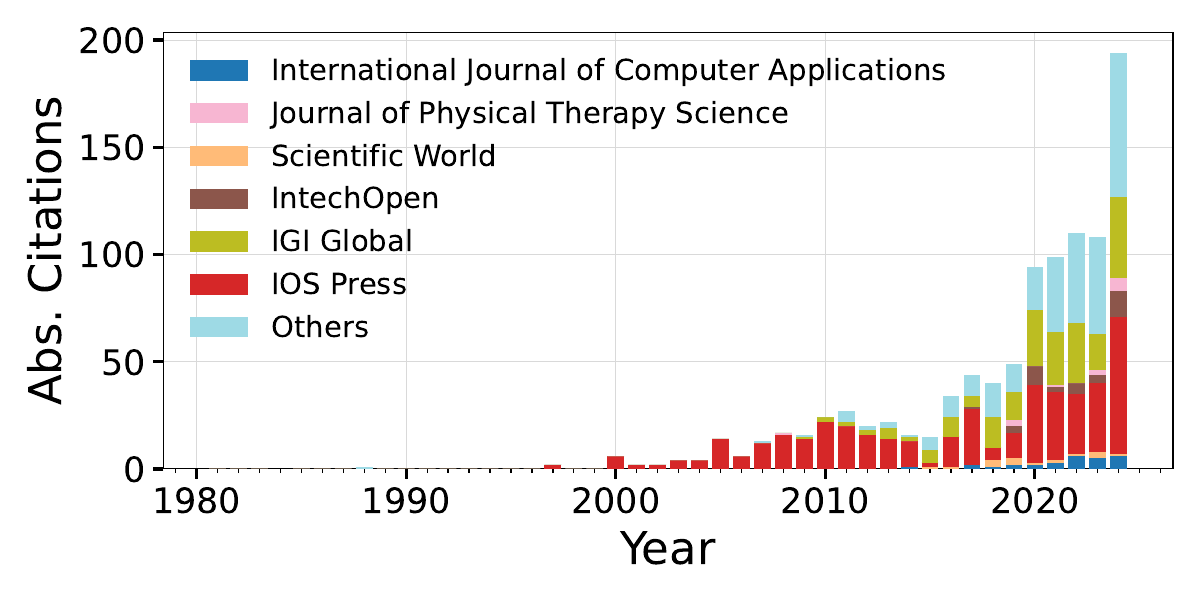}%
\captionsetup{labelformat=empty}
\caption{\rev{(c) Absolute number of citations to potential predatory journals and publishers (excluding Frontiers journals)}}
  \label{fig:beall-c}%
\end{subfigure}
\hfill
\begin{subfigure}[b]{0.32\textwidth}
     \centering
  \includegraphics[width=\textwidth]{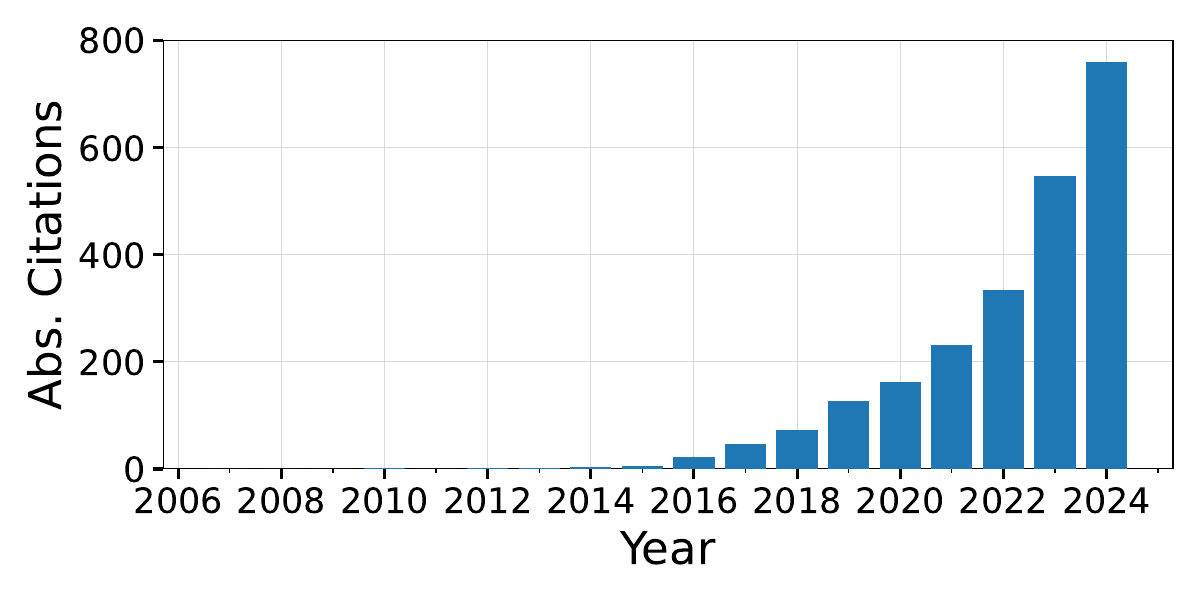}%
\captionsetup{labelformat=empty}
\caption{\rev{(e) Absolute number of citations to Frontiers journals from CHI articles\\}}
  \label{fig:beall-e}%
\end{subfigure}
\\[.4\baselineskip]
\begin{subfigure}[b]{0.32\textwidth}
     \centering
  \includegraphics[width=\textwidth]{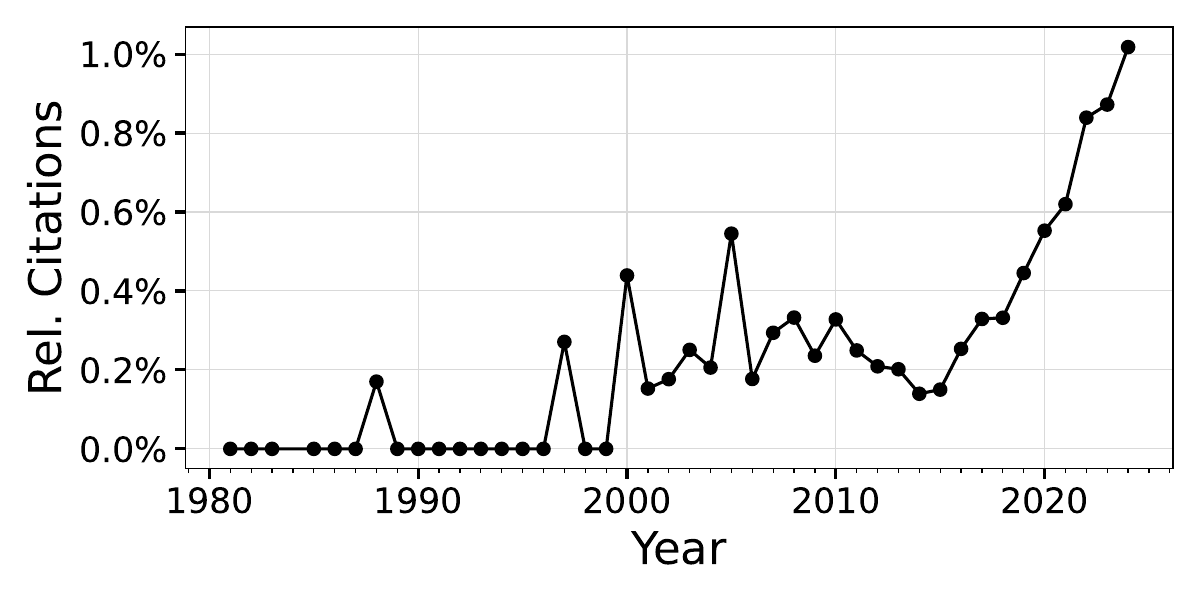}%
\captionsetup{labelformat=empty}
  \caption{(b) Relative number of citations to potential predatory journals and publishers\\}
  \label{fig:beall-b}%
\end{subfigure}
\hfill
\begin{subfigure}[b]{0.32\textwidth}
     \centering
  \includegraphics[width=\textwidth]{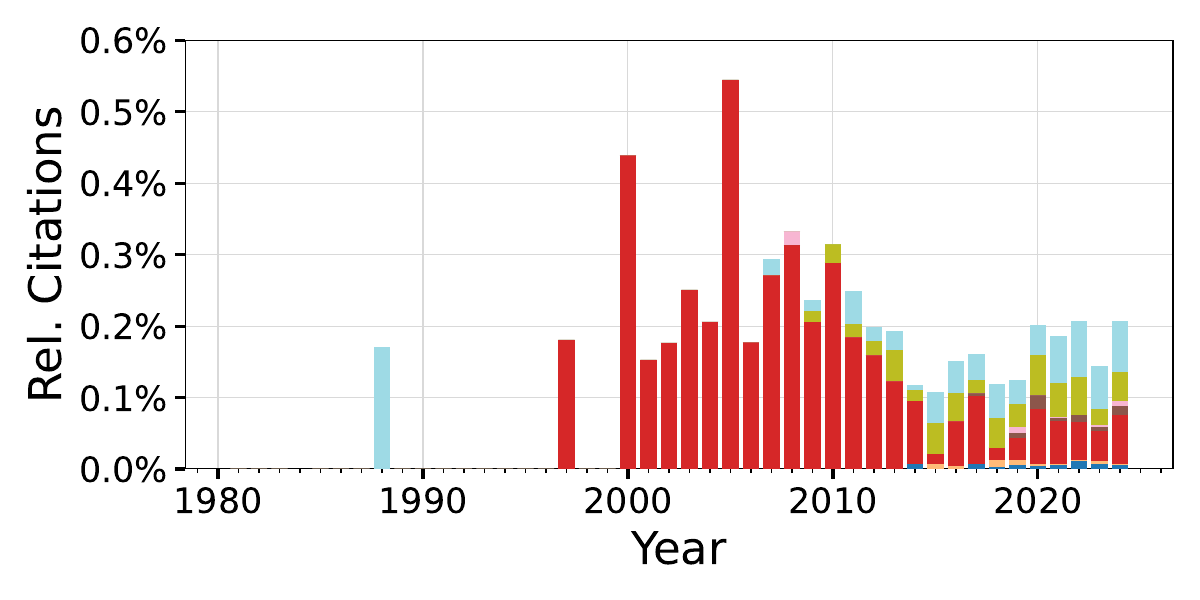}%
\captionsetup{labelformat=empty}
\caption{(d) Relative number of citations to potential predatory journals and publishers (excluding Frontiers journals)}
  \label{fig:beall-d}%
\end{subfigure}
\hfill
\begin{subfigure}[b]{0.32\textwidth}
     \centering
  \includegraphics[width=\textwidth]{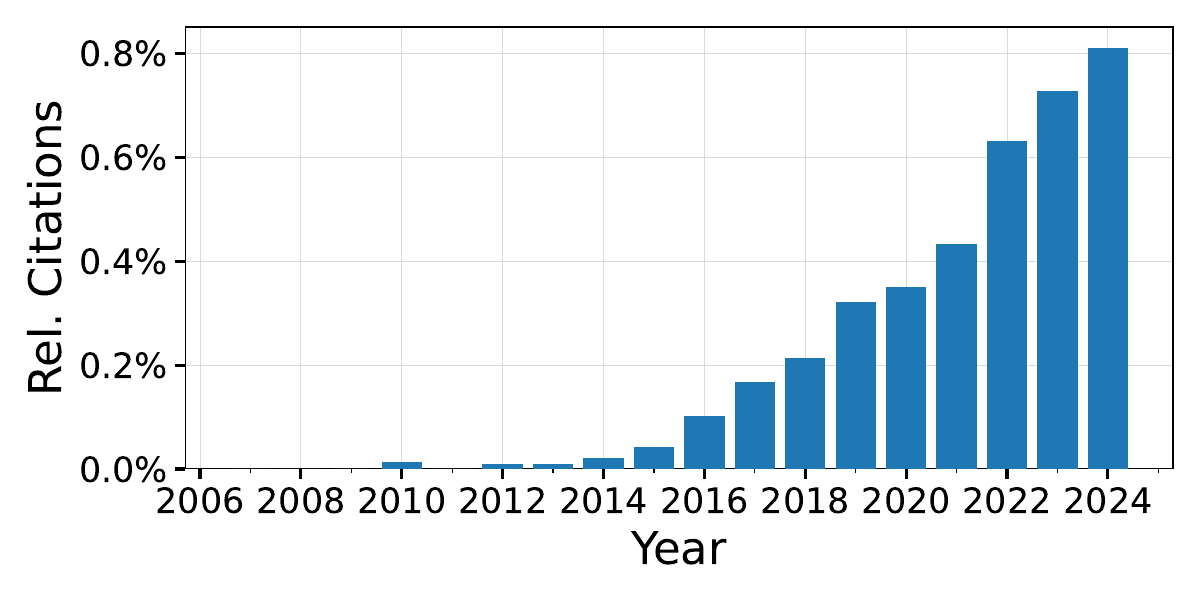}%
\captionsetup{labelformat=empty}
\caption{(f) Relative number of citations to Frontiers journals from CHI articles\\}
  \label{fig:beall-f}%
\end{subfigure}
  \caption{Citations to potential predatory journals and publishers from CHI articles have become more common since 2016 (a and b).
    This growth is to a great extent produced by citations to large questionable publishers, such as Frontiers Media (e and f).
    While citations from CHI articles to smaller questionable publishers are growing in absolute terms (c), they remain limited in the recent decade in relative terms (d).
  % Left: a) absolute and d) relative number of citations to potential predatory journals and publishers in Beall's List;
  % Middle: b) absolute and e) relative citations to potential predatory journals and publishers other than Frontiers journals;
  % Right: c) absolute and f) relative citations to Frontiers journals.
  }%
  \label{fig:beall}%
\end{figure*}%
%TC:endignore

%% file: TAB-EVENTSTUDY.tex
%TC:ignore
\begin{table*}[!thbp]%
\caption{The results of an event study indicate a significant change in the mean number of references (dependent variable) per CHI article before (2007--2015) and after (2016--2024) the decision to lift page restrictions at the CHI Conference.
}%
\label{tab:eventstudy}%
% \small
\begin{tabular}{lrrrrr}%
\toprule
     Statistic & Value &  &  &  &   \\
\midrule
    R\textsuperscript{2} &
        0.999
    & \multicolumn{2}{l}{df\textsubscript{regression}}
        & 9\hphantom{*}
    \\
    Adj. R\textsuperscript{2} &
        0.997
    & \multicolumn{2}{l}{df\textsubscript{residuals}}
        & 8\hphantom{*}
    \\
    F-test &
    $F=687.6$, $p<1e^{-9}$
    \\
\midrule
     Parameter & Coefficient & std err & t & p\hphantom{*} & CI (95\%)  \\
\midrule
    % Year range (before) & 2007--2015 \\
    % Year range (after)  & 2016--2024 \\
    % Regression equation &
    % \small
    % $y={\beta_0} + {\beta_1}{\mathrm{t}}$ % + {\epsilon_t}$
    % \\
    Coefficient $\beta_0$ (intercept) &
                34.9467   &   2.498 &    13.992   &   0.000*  &    [29.187,      40.706]
                    \\
    Coefficient $\beta_1$ (indicator) &
                  6.5244    &  1.771    &  3.683    &  0.006*    &   [2.439,      10.609]
                       \\
    Coefficient $\beta_2$ (AwardRatio) &
                1.9100  &    5.631   &   0.339  &    0.743\hphantom{*}   &  [-11.075,      14.895]
                  \\
    Coefficient $\beta_3$ (Authors) &
              -9.2715   &   5.969  &   -1.553   &   0.159\hphantom{*}   &  [-23.035,       4.492]
                  \\
    Coefficient $\beta_4$ (Arxiv) &
                -3.6121   &   1.651  &   -2.188    &  0.060\hphantom{*}  &    [-7.419,       0.195]
                      \\
    Coefficient $\beta_5$ (Repos) &
                -2.7982  &   29.674 &    -0.094   &   0.927\hphantom{*}   &  [-71.227,      65.631]
                       \\
    Coefficient $\beta_6$ (Reviews) &
              12.8483  &   48.702  &    0.264   &   0.799\hphantom{*}  &   [-99.459,     125.156]
                     \\
    Coefficient $\beta_7$ (Predators) &
            38.7569   &  10.727   &   3.613    &  0.007*   &   [14.021,      63.493]
                   \\
    Coefficient $\beta_8$ (Interaction) &
     3.3381   &   0.999   &   3.341    &  0.010*    &   [1.034,       5.642]
               \\
    Coefficient $\beta_t$ (Time) &
         1.8930  &    0.535  &    3.537   &   0.008*   &    [0.659,       3.127]
                \\
% \midrule
%      Omnibus & 1.415, $p = 0.493$ &  &  &  &   \\
%      Skew & 1.415 &  &  &  &   \\
%      Omnibus & -0.385 &  &  &  &   \\
%      Kurtosis & 2.069 &  &  &  &   \\
%      Durbin-Watson & 1.680 \\
%      Jarque-Bera (JB) & 1.095, $p=0.578$ \\
%      Cond. No. & 2.43e+04 \\
\bottomrule
\end{tabular}%
\end{table*}%
%TC:endignore